\documentclass[letterpaper]{article}\usepackage[]{graphicx}\usepackage[]{color}
\pdfoutput=1

\makeatletter
\def\maxwidth{ %
  \ifdim\Gin@nat@width>\linewidth
    \linewidth
  \else
    \Gin@nat@width
  \fi
}
\makeatother

\usepackage{Sweave}

\usepackage[]{graphicx}\usepackage[]{color}
\pdfoutput=1
\makeatletter
\def\maxwidth{ %
  \ifdim\Gin@nat@width>\linewidth
    \linewidth
  \else
    \Gin@nat@width
  \fi
}
\makeatother

\usepackage[letterpaper, total={6.5in, 9in}]{geometry}

\usepackage[]{graphicx}\usepackage[]{color}
\makeatletter
\def\maxwidth{ %
  \ifdim\Gin@nat@width>\linewidth
    \linewidth
  \else
    \Gin@nat@width
  \fi
}

\let\proglang=\textsf
\newcommand{\pkg}[1]{{\fontseries{b}\selectfont #1}}
\let\url=\texttt

\newcommand*{\affaddr}[1]{\small{#1}} 
\newcommand*{\affmark}[1][*]{\textsuperscript{#1}}

\newcommand{\code}[1]{\mbox{\texttt{#1}}}
\usepackage[authoryear,round]{natbib}
\bibpunct{(}{)}{;}{a}{}{,}
\usepackage{algorithm}
\usepackage{algpseudocode}
\usepackage{pifont}
\usepackage{amsmath, amsthm, amssymb}
\usepackage{subfig}
\usepackage{natbib}
\IfFileExists{upquote.sty}{\usepackage{upquote}}{}
\IfFileExists{upquote.sty}{\usepackage{upquote}}{}
\IfFileExists{upquote.sty}{\usepackage{upquote}}{}
\begin{document}

  \title{Sequential Monte Carlo Methods in the \pkg{nimble} R Package}
\date{}
\author{%
Nicholas Michaud\affmark[1]\affmark[2], Perry de Valpine\affmark[2], Daniel Turek\affmark[3],\\
 Christopher J. Paciorek\affmark[1], Dao Nguyen\affmark[4]\\
\affaddr{\affmark[1]Department of Statistics, University of California, Berkeley}\\
\small{\affmark[2]Department of Environmental Science, Policy, and Management,}\\
 \small{University of California, Berkeley}\\
\affaddr{\affmark[3] Department of Mathematics and Statistics, Williams College}\\
\affaddr{\affmark[4] Department of Mathematics, University of Mississippi}
}

\maketitle
 \begin{abstract}  
 \pkg{nimble} is an \proglang{R} package for constructing algorithms and conducting inference on hierarchical models.  The \pkg{nimble} package provides a unique combination of flexible model specification and the ability to program model-generic algorithms. Specifically, the package allows users to code models in the \proglang{BUGS} language, and it allows users to write algorithms that can be applied to any appropriate model.  In this paper, we introduce \pkg{nimble}'s capabilities for state-space model analysis using sequential Monte Carlo (SMC) techniques.  We first provide an overview of state-space models and commonly-used SMC algorithms.  We then describe how to build a state-space model and conduct inference using existing SMC algorithms within \pkg{nimble}.  SMC algorithms within \pkg{nimble} currently include the bootstrap filter, auxiliary particle filter, ensemble Kalman filter, IF2 method of iterated filtering, and a particle MCMC sampler.   These algorithms can be run in \proglang{R} or compiled into \proglang{C++} for more efficient execution.  Examples of applying SMC algorithms to linear autoregressive models and a stochastic volatility model are provided.  Finally, we give an overview of how model-generic algorithms are coded within \pkg{nimble} by providing code for a simple SMC algorithm.  This illustrates how users can easily extend nimble's SMC methods in high-level code.
\end{abstract}

  \section[Introduction]{Introduction}

  State-space models provide a framework for analyzing time-series data, where observations are assumed to be noisy measurements of unobserved latent states that evolve over time \citep{durbin12}.  State-space models have been used in such diverse fields as population ecology \citep{knape12}, epidemiology \citep{andersson12}, economics \citep{fernandez07}, and meteorology \citep{wikle13}.  With the broad applicability of state-space models has come a variety of methods for conducting inference.  One common goal is to determine the filtering distribution of the model, that is, the distribution of the most recent latent states given data.  A second goal lies in estimating the likelihood of the model by integrating over the latent state dimensions, which can in turn be used to obtain maximum likelihood or Bayesian estimates of top-level parameters.  State-space models are also frequently used for forecasting and for estimating the smoothing distribution, which is the distribution of all previous latent states given the data.
  
  For linear, Gaussian state-space models, the Kalman filter gives an exact solution to the above problems.  However, for more general state-space models, analytical solutions are usually unavailable.  Modifications such as the extended Kalman filter \citep{anderson79}, which uses linearization to approximate the filtering distribution, have been applied to non-linear, non-Gaussian models but can yield inaccurate results.   More recently, it has become common to use a set of flexible computational algorithms known as sequential Monte Carlo (SMC) methods \citep{doucet01}.  
  
  SMC methods, or variations thereof, are attractive as they provide a general framework for conducting inference on any state-space model.  SMC algorithms estimate the filtering distribution using sequential importance sampling (SIR) \citep{doucet01}, although different versions differ in their details.  Generally speaking, SMC algorithms proceed by tracking a set of particles through time, where each particle represents a sample from the filtering distribution of the latent state.  When a new data point is received, the particles are updated via SIR to represent the filtering distribution given the most current data. In this manner, SMC methods can be used to perform ``on-line'' inference.  ``On-line'' inference is inference on filtering and forecasting distributions that can be updated iteratively as more data are received, without the need to re-process previously-received data.  In contrast, ``off-line'' methods such as MCMC must be re-run with the entire set of data each time a new data point is obtained.  SMC methods are nevertheless also of interest in ``off-line'' problems when MCMC or other approaches present difficulties.
  
 A variety of SMC methods currently exist, including the bootstrap filter \citep{gordon93}, auxiliary particle filter \citep{pitt99}, Liu-West filter \citep{liu01}, Storvik filter \citep{storvik02}, particle learning algorithm \citep{carvalho10}, ensemble Kalman filter \citep{evensen03}, and others.  In addition, algorithms such as particle MCMC \citep[PMCMC; ][]{andrieu10} have been developed that place SMC methods within a broader MCMC framework.  SMC algorithms are also a foundation for maximum likelihood estimation such as by the IF2 version of iterated filtering \citep{ionides15}.  SMC algorithms have also been used to conduct approximate Bayesian computation \citep{delmoral12}.   
  
 The generality of SMC methods makes them well-suited for implementation within the \pkg{nimble} \proglang{R} software package \citep{devalpine16}.  \pkg{nimble} adopts and extends a dialect of the hierarchical modeling language used by \proglang{WinBUGS}, \proglang{OpenBUGS} and \proglang{JAGS} \citep{lunn00, lunn12, plummer03}.  These models can then be analyzed using \pkg{nimble}'s library of model-generic algorithms.  Additionally, \pkg{nimble} provides a programming system embedded in \proglang{R} for users to write their own model-generic algorithms.  These algorithms can be run in \proglang{R} or compiled  via automatically generated \proglang{C++} for more efficient execution.  In this paper,  \pkg{nimble}'s  SMC algorithms are described in detail.  Because they are all written in \pkg{nimble}'s algorithm programming system, they can be readily inspected and modified by users, much the way many base \proglang{R} functions are written in \proglang{R}.  \pkg{nimble} also has a variety of MCMC algorithms for more general Bayesian inference, as well as an MCEM algorithm.  
  
  In Section~\ref{sec:nimcompare}, we provide a more detailed comparison between \pkg{nimble} and other packages that can be used for state-space and more general hierarchical modeling. In Section~\ref{sec:ssms}, we introduce state-space models and the idea of filtering distributions.  We then describe a variety of algorithms that can be used for inference on filtering distributions.  In Section~\ref{subsec:nimbleModels}, we provide examples of writing state-space models within \pkg{nimble}.  Section~\ref{subsec:fixedParam} demonstrates inference for a linear autoregressive state-space model with fixed parameters. Section~\ref{subsec:unknownParam} gives examples of inference on top-level parameters using PMCMC for a stochastic volatility model and using  IF2  for a linear random-walk state-space model.  In both cases, the examples enable verification of correct behavior as well as exploration of sensitivity to key tuning parameters of each method.  In Section~\ref{sec:nimbleFuncs}, we demonstrate \pkg{nimble}'s programmability by coding an SMC algorithm within \pkg{nimble}'s algorithm programming system.

\section[nimble in comparison to other software packages]{nimble in comparison to other software packages}
\label{sec:nimcompare}

\pkg{nimble} can be used to conduct inference on hierarchical models via built-in SMC methods, MCMC methods, and other applicable built-in algorithms such as ascent-based Monte Carlo expectation-maximization \citep{caffo05}.  \pkg{nimble} can also be used as a tool that allows users to write functions in an \proglang{R}-like domain-specific language for model-generic algorithms, which we refer to as its algorithm programming system.  These can then be compiled and run in \proglang{C++} for increased speed.  In Section~\ref{sec:nimbleFuncs} we demonstrate creating a user-defined SMC algorithm written in \pkg{nimble}'s algorithm programming system.

\pkg{nimble}  offers a suite of sequential Monte Carlo algorithms that can be applied to state-space models written in the \proglang{BUGS} language  \citep{lunn12}. \pkg{Biips}, which has an \proglang{R} interface \pkg{rbiips} \citep{todeschini14}, similarly allows inference via SMC to be performed on \proglang{BUGS} models.  Compared to \pkg{nimble}, \pkg{Biips} acts as a ``black-box'' tool for SMC inference that automatically chooses an SMC algorithm for a given \proglang{BUGS} model.  \pkg{nimble} allows users to choose and customize the SMC method they would like to use.   Other current software packages that implement SMC algorithms include the \pkg{pomp} \proglang{R} package \citep{king16}, the \pkg{LibBi} software \citep{murray15} (with interface to \proglang{R} via the RBi package \citep{jacob18}), and the \pkg{vSMTC} \proglang{C++} template library \citep{zhou15}.  Of these, both \pkg{pomp} and \pkg{LibBi} offer algorithms not found in \pkg{nimble}, with \pkg{pomp} allowing for approximate Bayesian computation \citep{toni09}, and \pkg{LibBi} providing an SMC$^{2}$ algorithm \citep{chopin11}.
\pkg{nimble}, on the other hand, allows users to write their own SMC algorithms using \pkg{nimble}'s algorithm programming system.  Additionally, state-space models in \pkg{nimble} can benefit from \pkg{nimble}'s other algorithms.  For example, state-space models created in \pkg{nimble} can be compared to each other using \pkg{nimble}'s WAIC or cross-validation metrics. 

For linear, Gaussian state-space models, SMC algorithms are unnecessary, as exact inference can be conducted via the Kalman filter \citep{kalman60}.  \proglang{R} packages that implement the Kalman filter include  \pkg{dlm} \citep{petris10} and \pkg{dse} \citep{gilbert06}.  The \pkg{KFAS} \proglang{R} package \citep{helske17} allows for variations of Kalman filter inference to be conducted on models with non-Gaussian observation equations.  See \citet{petris11} and \citet{tusell11} for a detailed comparison of these packages.

The \pkg{TMB} \proglang{R} package enables maximum likelihood estimation of state-space models via Laplace approximation \citep{kristensen16}, using the \pkg{CppAD} \proglang{C++} package to enable automatic differentiation of likelihood functions \citep{bell05}. To conduct inference in \pkg{TMB}, a user must write a \proglang{C++} function that returns the log-likelihood of the state-space model.  In contrast, \pkg{nimble} currently does not support Laplace approximation, although maximum likelihood estimation for state-space models can be achieved through the IF2 algorithm described in Section~\ref{subsubsec:IF2}.

\pkg{nimble} also allows MCMC algorithms to be be flexibly composed and controlled by users.  The ability to try such different methods on the same models is distinct from other package, such as \pkg{WinBUGS} and \pkg{OpenBUGS} \citep{lunn00, lunn12}, \pkg{JAGS} \citep{plummer03}, and \pkg{Stan} \citep{carpenter17}.  It means, for example, that particle MCMC algorithms in \pkg{nimble} can harness \pkg{nimble}'s MCMC implementations.

\section[Sequential Monte Carlo methods for state-space models]{Sequential Monte Carlo methods for state-space models}  \label{sec:ssms}

\subsection{State-space models}

State-space models, also known as hidden Markov models, are used to model time-series data or any sequential data.  The vector of data at each time $t$, labeled ${Y}_{t}$, is assumed to be related to a latent, unobserved state ${X}_{t}$ through an observation distribution ${Y}_{t}\sim g_{t}({y}_{t}|{x}_{t}, {\theta})$.  Here, ${\theta}$ is a vector of top-level parameters that are assumed not to change with time.  ${X}_{t}$ depends on ${X}_{t-1}$ through a transition distribution ${X}_{t}\sim f_{t}({x}_{t}|{x}_{t-1},{\theta})$.   Frequently, the observation and transition distributions remain constant over all time points, in which case the $t$ subscripts on $f_{t}$ and $g_{t}$ are dropped. (Below, we drop the subscripts to simplify notation, but neither the algorithms nor \pkg{nimble}'s implementations require the distributions to remain constant.)  State-space models have the following Markov property: $f({x}_{t}|{x}_{1:t-1}, \theta) = f({x}_{t}|{x}_{t-1}, \theta)$,  where ${x}_{1:t-1}=\left(x_{1},\ldots,x_{t-1}\right)$.  Note that we assume no observation exists for $t=0$, and that ${x}_{0}$ comes from a known prior distribution $f({x}_{0}|{\theta})$. 

One goal is to determine the distribution $p({x}_{t}|{y}_{1:t},{\theta})$, known as the filtering distribution for ${X_{t}}$.   Consider a situation where new data are received sequentially in time, and data are currently available up to time $t-1$, that is, ${y}_{1:t-1}$ are known.  Upon receiving ${y}_{t}$, the filtering distribution $p(x_{t}|y_{1:t},\theta)$ provides information about the latent state at the most recent time point, given all data.  Similarly, the smoothing distribution $p(x_{1:t}|y_{1:t},\theta)$ provides information about latent states from all time points given the most recent data.  The forecast distribution $p(x_{t+\tau}|y_{1:t},\theta)$ for $\tau > 0$ is also frequently of interest.  Another common goal is to calculate the likelihood $p(y_{1:t}|\theta)$.  This likelihood can in turn be used to obtain maximum likelihood estimates of $\theta$, or in an MCMC framework to obtain samples from the posterior $p({\theta}|{y}_{1:t})$. 

SMC methods, at each time $t$, keep track of $M$ weighted samples $\left\{x_{t}^{(m)}\right\}_{m=1}^{M}$ from the filtering distribution $p(x_{t}|y_{1:t}, \theta)$, along with associated weights $\left\{\pi_{t}^{(m)}\right\}_{m=1}^{M}$.  These samples are known as ``particles''.  When a new data point $y_{t+1}$ is considered, a combination of propagating particles forward, re-weighting and/or re-sampling them is done to obtain an updated sample  $p(x_{t+1}|y_{1:t+1}, \theta)$ with updated weights  $\left\{\pi_{t+1}^{(m)}\right\}_{m=1}^{M}$.  The steps involved also provide an approximation of $p(y_{t+1} | y_{1:t}, \theta)$.  The methods differ in how the propagation, re-weighting and/or re-sampling are done.
  
As compared to MCMC methods, SMC methods generally perform well at generating samples, given fixed $\theta$, from the filtering, smoothing, and forecasting distributions, and at estimating model likelihoods.  SMC methods are also able to perform ``on-line'' inference.  On the other hand, SMC methods alone do not provide an easy way to achieve inference on top-level parameters.  Recently, algorithms such as PMCMC (described in Section~\ref{subsubsec:PMCMC}) and IF2 (described in Section~\ref{subsubsec:IF2}) have been proposed that use SMC methods in conjunction with MCMC and maximum likelihood algorithms, respectively, to achieve top-level parameter inference.

 MCMC algorithms can also be used to draw samples from the filtering, smoothing, or forecasting distributions, as well as samples of top-level parameters conditioned on the current data.  However, MCMC algorithms can encounter difficulty in producing valid proposals for ${x}_{1:t}$ due to the often-high degree of correlation among the latent states at each time point.  Attempts have been made at creating proposal distributions that account for these correlations, as in \citet{pooley15} and \citet{newman08}, but applying generic MCMC algorithms to state-space models does not always work efficiently.

\subsection{Filtering algorithms} 
\label{subsec:Filtering}

  In Section~\ref{subsubsec:Boot} and Section~\ref{subsubsec:Aux}, two types of SMC methods (the bootstrap filter and auxiliary particle filter) are described, each of which can be used to generate samples from the filtering distribution $p(x_{t}|y_{1:t}, \theta)$ or the smoothing distribution $p(x_{1:t}|y_{1:t},\theta)$, and to obtain likelihood estimates.  In Section~\ref{subsubsec:IF2}, we describe the IF2 algorithm, which uses SMC algorithms to produce maximum-likelihood estimates of $\theta$.  In Section~\ref{subsubsec:PMCMC}, a PMCMC algorithm is detailed that uses SMC methods to estimate likelihoods within a Metropolis-Hastings MCMC sampling scheme for ${\theta}$.
  
  The Ensemble Kalman filter, or EnKF \citep{evensen03}, can also be used to estimate the filtering distribution.  Similar to other SMC techniques, the EnKF approximates the filtering distribution via a collection of particles that are propagated forwards in time.  However, whereas other SMC methods use importance sampling to select particles at each time point, the EnKF instead shifts particles towards the filtering distribution using an approximation to the Kalman gain matrix.  The EnKF is described in Section~\ref{subsubsec:ENKF}.  
  
  The algorithm summaries provided in this paper are not intended to be a comprehensive overview of the current state of filtering methods.  Several papers describe the landscape of filtering algorithms and related inference methods for state-space models, including  \citet{arulampalam02}, \citet{doucet09}, and \citet{kantas15}.  

Note that in the following discussions of the bootstrap, auxiliary particle and ensemble Kalman filters, $\theta$ is treated as fixed, so we have omitted it to reduce notational overhead.

\subsection{Bootstrap filter}
\label{subsubsec:Boot}

The bootstrap filter of \citet{gordon93} uses importance sampling to sequentially generate samples from $p(x_{t}|y_{1:t})$ and approximate $p(y_t | y_{1:t-1})$ at each time $t$.  The bootstrap filter first propagates each particle from time $t-1$ forward according to a proposal distribution $\tilde{x}_{t}^{(m)}\sim q(x_{t}|x_{t-1}^{(m)},y_{t})$. Importance weights $\pi_{t}^{(m)}$ are then calculated for each particle.  In the basic version of the algorithm, the propagated particles are resampled according to these weights at each step.  This results in an equally weighted sample $\left\{x_{t}^{(m)}\right\}_{m=1}^{M}$ from $p(x_{t}|y_{1:t})$.  (See below for extensions that do not resample at each step.)
     
\begin{algorithm}[h]
	\caption{Bootstrap filter}
	\label{bootstrap_algo}
  \begin{algorithmic}[1]
    \For{$m$ in $1:M$}
    \State Generate $x_{0}^{(m)}\sim f(x_{0})$\label{line:init1}
    \State Set $\pi^{(m)}_{0} = \frac{1}{M}$\label{line:init2}
    \EndFor 
		\For{$t$ in $1:T$}
		\For{$m$ in $1:M$} \label{line:bf-kloop}
		\State Generate $\tilde{x}_{t}^{(m)}\sim q({x}_{t}|x_{t-1}^{(m)},y_{t})$\label{line:xsamp}
		\State Calculate unnormalized weight $w_{t}^{(m)} = \frac{f(\tilde{x}_{t}^{(m)}|x_{t-1}^{(m)})g(y_{t}|\tilde{x}_{t}^{(m)})}{q(\tilde{x}_{t}^{(m)}|x_{t-1}^{(m)},y_{t})}\pi_{t-1}^{(m)}$\label{line:bootWeight}
		\EndFor
		\For{$m$ in $1:M$}
		\State Normalize $w_{t}^{(m)}$ as $\pi_{t}^{(m)} = \frac{w_{t}^{(m)}}{\sum_{i=1}^{M}w_{t}^{(i)}}$.
 		\EndFor
		\For{$m$ in $1:M$}
		\State Sample an index $j_{m}$ from the set {1,...,M} with probabilities $\{\pi_{t}^{(m)}\}_{m=1}^{M}$. \label{line:ewsamp1}
		\State Set $x_{t}^{(m)} = \tilde{x}_{t}^{(j_{m})}$. \label{line:ewsamp2}
		\State Set $\pi_{t}^{(m)}=\frac{1}{M}$\label{line:eweight}
		\EndFor 
    \State Calculate $\tilde{p}(y_{t|1:t-1}) = \frac{1}{M}\sum_{m=1}^{M}w_{t}^{(m)}$\label{line:bootLikelihood}
		\EndFor
	\end{algorithmic}
\end{algorithm}

\pkg{nimble} uses a simple choice for $q( \cdot | \cdot)$, namely the forward simulation density, $f(x_{t}^{(m)}|x_{t-1}^{(m)})$.  With this choice, Step 8 simplifies to $w_{t}^{(m)} = g(y_{t}|\tilde{x}_{t}^{(m)}) \pi_{t-1}^{(m)}$.  Then one does not need to calculate $f(\tilde{x}_{t}^{(m)}|x_{t-1}^{(m)})$, which is important because these methods are sometimes used when it is feasible to simulate forward in time but not actually calculate the forward density.   Note that resampling (Steps~\ref{line:ewsamp1}-\ref{line:eweight} in Algorithm~\ref{bootstrap_algo}) creates an equally weighted sample from the target distribution of latent states at time $t$.  Additionally, an estimate of the likelihood $p(y_{1:T})$ can be obtained by $\tilde{p}(y_{1:T})=\prod_{t=1}^{T}\tilde{p}(y_{t|1:t-1})$, where $\tilde{p}(y_{t|1:t-1})$ is given in line~\ref{line:bootLikelihood} of the algorithm.

The resampling step of Algorithm~\ref{bootstrap_algo} is performed to reduce particle degeneracy.  Particle degeneracy is a common problem in filtering algorithms, where a small number of particles have most of the weight placed on them, leaving most particles with low weights \citep{doucet00}. Particle degeneracy corresponds to high Monte Carlo variance of approximations made using the filtered particles.  It causes the filter to spend computational effort in propagating and weighting particles that contribute little to our knowledge of the target distribution.  Resampling ensures that mostly highly-weighted particles will be propagated forwards, increasing algorithm efficiency by providing a better estimate of the target distribution and likelihood. 

However, resampling particles at each time point can also lead to a loss of particle ``diversity'' \citep{doucet00}, as many of the resampled particles at each time point will have the same value.  Thus it has been proposed \citep{smith01} that resampling should take place only if particle degeneracy becomes too significant.  An estimate of particle degeneracy is the effective sample size, calculated at each time $t$ as $ESS = \frac{1}{\sum_{m=1}^{M}\left(\pi_{t}^{(m)}\right)^{2}}$.  \citet{smith01} recommend that resampling should be conducted only if the effective sample size becomes too low, indicating many particles with low weights.  As a criterion for when a resampling step should take place, a threshold $\tau$ is chosen with $0\leq \tau \leq 1$, such that the algorithm will resample particles whenever $\frac{ESS}{M} < \tau$.  Note that choosing $\tau = 0$ will mean that the resampling step is never performed, and choosing $\tau=1$ will ensure that sampling is performed at each time point.  To perform the above algorithm without resampling, simply remove Steps~\ref{line:ewsamp1}-\ref{line:eweight}.  If the resampling step is not performed, the set $\left\{\tilde{x}_{t}^{(m)}, \pi_{t}^{(m)} \right\}_{m=1}^M$ will constitute an unequally weighted sample from the target distribution.

  Various methods for resampling particles have been employed, including systematic resampling, stratified resampling, residual resampling, and multinomial resampling \citep{doucet09}.  Generally, systematic resampling, stratified resampling, and residual resampling have been found to perform similarly, and to outperform multinomial resampling due to its higher variance \citep{douc05}.  \pkg{nimble} allows users to choose from any of these four methods.  Systematic resampling is used by default.
  
  Additionally, the above filter can be used to produce samples from the smoothing distribution $p(x_{1:t}|y_{1:t})$ using methods described in \citet{doucet09}.  \pkg{nimble} currently uses the naive approach of storing the ancestors of the set of particles at time $t$.  Specifically, using the notation of \citet{andrieu10}, \pkg{nimble}'s smoothing method keeps track of the set $x_{1:T}^{(m)} = \left(x_{1}^{(B_{1}^{(m)})},  x_{2}^{(B_{2}^{(m)})}, \ldots, x_{T}^{(B_{T}^{(m)})}\right)$ for $m = 1, \ldots, M$, where $B_{t}^{(m)}$ is the index of the ancestor particle at time $t$ that gave rise to particle $m$ at time $T$.  We note that this approach to estimating the smoothing distribution can produce poor estimates of $p(x_{t}|y_{1:T}, \theta)$ when $T$ is much greater than $t$, as the number of unique ancestors at time $t$ will decrease as $T$ increases.  More accurate methods for estimating the smoothing distribution, such as forward-backward smoothing, can be found in \citet{kantas15}.
  
  Though the bootstrap filtering algorithm was first put forth in 1993, it remains commonly used in a wide variety of applications, such as inferring the distribution of radii of planets \citep{silburt15} and estimating neurological activity \citep{croce17}.  \citet{yang13} use a bootstrap filter to help to normalize text obtained from social media. \citet{oladyshkin13} employ a bootstrap filter to model CO$_{2}$ storage in geological formations.

\subsection{Auxiliary particle filter}
\label{subsubsec:Aux}
The auxiliary particle filter algorithm (APF) of \citet{pitt99} uses importance sampling similarly to the bootstrap filter but includes an additional ``look-ahead step''.  At each time $t$, the APF calculates first-stage weights $w_{t|t-1}^{(m)}$ for particles from time $t-1$.  These weights are calculated using a rough estimate of the likelihood of the current data given each particle from the previous time point, labeled $\hat{p}(y_{t}|x_{t-1}^{(m)})$.  Particles with high first-stage weights correspond to values of the latent state at time $t-1$ that are likely to generate the observed data at time $t$.  To estimate $\hat{p}(y_{t}|x_{t-1}^{(m)})$, \citet{pitt99} recommend choosing an auxiliary variable $\tilde{x}^{(m)}_{t|t-1}$ and then setting $\hat{p}(y_{t}|x_{t-1}^{(m)})=p(y_{t}|\tilde{x}^{(m)}_{t|t-1})$.  Possible methods for choosing  $\tilde{x}^{(m)}_{t|t-1}$ include simulating a value from $f(x_{t}|x_{t-1}^{(m)})$, or taking $\tilde{x}^{(m)}_{t|t-1} = E(x_{t}|x_{t-1}^{(m)})$.  

The first-stage weights are used to sample $M$ particles from time $t-1$, labeled $\tilde{x}_{t-1}^{(m)}$ for $m=1,\ldots,M$.  The sampled particles are then propagated forwards by a proposal distribution $q(x_{t}^{(m)}|\tilde{x}_{t-1}^{(m)},y_{t})$ and reweighted using second-stage weights $w_{t}^{(m)}$, providing a weighted sample from $p(x_{t}|y_{1:t})$.  The APF as shown in \citet{pitt99} optionally includes a second resampling step using the second-stage weights.  However, the algorithm using a single resampling step has been shown to be more efficient \citep{carpenter99}.
\begin{algorithm}[ht] 
	\caption{Auxiliary particle filter}
	\label{apf_algo}
	\begin{algorithmic}[1]
   \For{$m$ in $1:M$}
    \State Generate $x_{0}^{(m)}\sim f(x_{0})$
    \State Set $\pi^{(m)}_{0} = \frac{1}{M}$
    \EndFor
		\For{$t$ in $1:T$}
		\For{$m$ in $1:M$}
                \State Generate $\tilde{x}^{(m)}_{t|t-1}$ from either $\tilde{x}^{(m)}_{t|t-1} \sim f(x_{t}|x_{t-1}^{(m)})$ or $\tilde{x}^{(m)}_{t|t-1} = E(x_{t}|x_{t-1}^{(m)})$\label{line:aux1stwtA}
                \State Calculate $\hat{p}(y_{t}|x_{t-1}^{(m)})=p(y_{t}|\tilde{x}^{(m)}_{t|t-1})$ 
		\State Calculate unnormalized weight $w_{t|t-1}^{(m)} = \pi_{t-1}^{(m)}\hat{p}(y_{t}|x_{t-1}^{(m)})$\label{line:aux1stwtB}
    \EndFor
		\For{$m$ in $1:M$}
		\State Normalize $w_{t|t-1}^{(m)}$ as $\pi_{t|t-1}^{(m)} = \frac{w_{t|t-1}^{(m)}}{\sum_{i=1}^{M}w_{t|t-1}^{(i)}}$
    \EndFor
		\For{$m$ in $1:M$}
		\State Sample an index $j_{m}$ from the set {1,...,M} with probabilities $\{\pi_{t|t-1}^{(m)}\}_{m=1}^{M}$.
		\State Set $\tilde{x}_{t-1}^{(m)} = x_{t-1}^{(j_k)}$
		\State Generate $x_{t}^{(m)} \sim q(x_{t}|\tilde{x}_{t-1}^{(m)}, y_{t})$
		\State Calculate unnormalized weight $w_{t}^{(m)} = \frac{f(x_{t}^{(m)}|\tilde{x}_{t-1}^{(m)})g(y_{t}|x_{t}^{(m)})}{\hat{p}(y_{t}|x_{t-1}^{(m)})q(x_{t}^{(m)}|\tilde{x}_{t-1}^{(m)},y_{t})}$\label{line:auxwt}
		\EndFor
		\For{$m$ in $1:M$}
		\State Normalize $w_{t}^{(m)}$ as $\pi_{t}^{(m)} = \frac{w_{t}^{(m)}}{\sum_{i=1}^{M}w_{t}^{(i)}}$\label{line:auxresamp}
		\EndFor
    \State Calculate $\tilde{p}(y_{t|1:t-1}) = \left(\frac{1}{M} \sum_{m=1}^{M}  w_{t}^{(m)} \right)\left(\sum_{m=1}^{M}w_{t|t-1}^{(m)}\right)$\label{line:auxLikelihood}
		\EndFor
	\end{algorithmic}
\end{algorithm}

\pkg{nimble} again uses the basic choice $f(x_{t} | \tilde{x}_{t-1}^{(m)})$ for $q(\cdot | \cdot)$, which simplifies the weight in step~\ref{line:auxwt} to $w_{t}^{(m)}  = g(y_{t}|x_{t}^{(m)}) / \hat{p}(y_{t}| x_{t-1}^{(m)})$.  In a manner similar to the bootstrap filter, the APF can be used to obtain an estimate of the likelihood $p(y_{1:T})$ as $\tilde{p}(y_{1:T})=\prod_{t=1}^{T}\tilde{p}(y_{t|1:t-1})$, where $\tilde{p}(y_{t|1:t-1})$ is given in line~\ref{line:auxLikelihood} of the APF algorithm.  \pkg{nimble} provides the two choices for generating the first stage weights (steps~\ref{line:aux1stwtA}-\ref{line:aux1stwtB}) described above: either via forward simulation of the latent state or, when available for the specific model, via the predicted latent state mean from each particle at time $t-1$.

Similar to the bootstrap filter, the auxiliary particle filter was developed some time ago, but still sees frequent use.  Recently, APF algorithms have been applied to problems in fields such as pedestrian navigation \citep{yu17} and battery life prediction \citep{liu11}.

\subsection{IF2 Algorithm}
\label{subsubsec:IF2}

Unlike the bootstrap and auxiliary particle filters, the IF2 algorithm \citep{ionides15} is designed for maximum likelihood estimation.  The IF2 algorithm  is a variant of the basic form of iterated filtering, proposed by \cite{ionides06} and \cite{ionides11}. The original iterated filtering algorithm was found to have favorable performance and theoretical properties as compared to the Liu-West method of estimating the posterior distribution $p(\theta|y_{1:T})$ \citep{liu01}.  The innovation of \cite{ionides06} is to perturb $\theta$ by random walk noise and to use particle filter (SMC) likelihood estimates to approximate derivatives of the log-likelihood function in order to produce maximum likelihood estimates. 

\cite{ionides15} modified the theory developed by \cite{ionides06} by using an iterated Bayes maximum a posteriori (MAP) estimate in place of gradient estimates of the perturbed parameters $\theta$
to optimize the MLE. Iterating in this way has some benefits:
\begin{enumerate}
	\item A theoretical foundation for this method can be obtained by convergence of the iterated Bayes map.
	\item Methods that are not based on local polynomial approximations to the likelihood surface can be advantageous when the likelihood surface has nonlinear ridges.
	\item Computational expense is reduced by removing the need for the computationally
	demanding gradient of the log likelihood.
\end{enumerate}

The IF2 algorithm proceeds by running a modified particle filter for $I$ iterations.  As in the Liu-West filter, parameters $\theta$ are also represented by particles that are weighted and resampled along with their associated latent states.  The perturbations to the parameters are generated at each time step of each particle filter iteration.  The perturbations follow a schedule of decreasing magnitude to yield convergence to the maximum likelihood parameters.  In what follows, our notation differs somewhat from that of \citet{ionides15} to be both more specific and more consistent with the other algorithms described here.

For iteration $i$, let $h_{t}(\theta | \Sigma_{i})$ for $t = 1, \ldots, T$ be a sequence of user-chosen densities that represent perturbations of $\theta$.  Indexing the density by $t$ (the particle filter time index) and its parameters by $i$ (the IF2 iteration index) is how \citet{ionides15} present the method, but the associated implementation in \pkg{pomp} \citep{king16} uses a scheme where the $h(\cdot)$ does not change -- it is a normal distribution -- but rather its parameters change in both $t$ and $i$.  Hence from here on we write this as $h(\theta | \Sigma_{i, t})$ for clarity.  Formally, this can be viewed as a special case of $h_{t}(\theta | \Sigma_{i})$.  Choice of $h$ and $\Sigma_{i, t}$ is described below.

IF2 iteration $i$ begins with a parameter swarm $\left\{\theta_{i, 0}^{(m)}\right\}_{m = 1}^{M}$ where  $\theta_{i, 0}^{(m)} \sim h(\theta | \theta_{i-1, T}^{(m)}, \Sigma_{i,0})$.  A bootstrap particle filter is then run, with a key modification:  at each time step $t$, before simulating, weighting and resampling, we perturb parameters by drawing $\tilde{\theta}_{i, t}^{(m)} \sim h(\theta | \theta_{i, t-1}^{(m)}, \Sigma_{i, t})$.  These parameter values are then included in weighting and resampling along with their corresponding latent state values.  

Once again, for the choice for forward proposals, $q(\cdot | \cdot)$ in step~\ref{if2:q}, \pkg{nimble} uses the basic choice of simulating from the model's latent state dynamics, $\tilde{x}_{t}^{(m)}\sim f(x_{t}|x_{t-1}^{(m)}, \tilde{\theta}_{i, t}^{(m)})$.  The weight in step~\ref{if2:wt} then simplifies to $w_{t}^{(m)} = g(y_{t} | \tilde{x}_{t}^{(m)}, \tilde{\theta}_{i, t}^{(m)})$.

After the $I^{\mbox{th}}$ iteration, the parameter samples $\left\{\theta_{I, T}^{(m)}\right\}_{m=1}^{M}$ are averaged to produce an estimate of the MLE of $\theta$.  In the IF2 algorithm, by analogy with simulated annealing, it has been shown
that, with an appropriate schedule for decreasing perturbation magnitudes, in the limit as the perturbations $h(\theta | \Sigma_{i, t})$ go to zero variance with mean $\theta$ as $i \rightarrow \infty$, the algorithm should reach the maximum likelihood solution.   Algorithm~\ref{IF2_algo} presents the IF2 algorithm as it is implemented in \pkg{nimble}. 

\begin{algorithm}[ht] 
	\caption{IF2}
	\label{IF2_algo}
	\begin{algorithmic}[1]
	\For{$m$ in $1:M$}
	\State Generate $\theta_{0, T}^{(m)} \sim h(\theta|\theta_{\mbox{init}}, \Sigma_{\mbox{init}})$
	\EndFor
	\For{$i$ in $1:I$}
	\For{$m$ in $1:M$}
		\State Generate $\theta_{i, 0}^{(m)} \sim h(\theta|\theta_{i-1, T}^{(m)}, \Sigma_{i, 0})$
		\State Generate $x_{0}^{(m)}\sim f(x_{0}|\theta_{i, 0}^{(m)})$
		\EndFor
		\For{$t$ in $1:T$}
		\For{$m$ in $1:M$}
		\State Generate $\tilde{\theta}_{i, t}^{(m)}\sim h(\theta | \theta_{i, t-1}^{(m)}, \Sigma_{i, t})$
		\State Generate $\tilde{x}_{t}^{(m)}\sim q({x}_{t}|x_{t-1}^{(m)},y_{t}, \tilde{\theta}_{i, t}^{(m)})$\label{if2:q}
		\State Calculate unnormalized weight $w_{t}^{(m)} = \frac{f(\tilde{x}_{t}^{(m)}|x_{t-1}^{(m)}, \tilde{\theta}_{i, t}^{(m)})g(y_{t}|\tilde{x}_{t}^{(m)}, \tilde{\theta}_{i, t}^{(m)})}{q(\tilde{x}_{t}^{(m)}|x_{t-1}^{(m)},y_{t}, \tilde{\theta}_{i, t}^{(m)})}$\label{if2:wt}
		\EndFor
		\For{$m$ in $1:M$}
		\State Normalize $w_{t}^{(m)}$ as $\pi_{t}^{(m)} = \frac{w_{t}^{(m)}}{\sum_{i=1}^{M}w_{t}^{(i)}}$ 
		\EndFor
		\For{$m$ in $1:M$}
		\State Sample an index $j_{m}$ from the set {1,...,M} with probabilities $\{\pi_{t}^{(m)}\}_{m=1}^{M}$.
		\State Set $x_{t}^{(m)} = \tilde{x}_{t}^{(j_{m})}$ and $\theta_{i, t}^{(m)} = \tilde{\theta}_{i, t}^{(j_{m})}$
		\EndFor
		\EndFor
		\EndFor 
		\State Return $\bar{\theta} = \frac{1}{M}\sum_{m=1}^{M}\theta_{I, T}^{(m)}$
	\end{algorithmic}
\end{algorithm}

\pkg{nimble}'s implementation of IF2, and specifically the schedule of decreasing (or cooling) perturbation magnitudes, follows that of \pkg{pomp} \citep{king16}.  $h(\theta | \theta_{i, t-1}^{(m)}, \Sigma_{i,t})$ is a product of independent normals, i.e., a multivariate normal with mean $\theta_{i, t-1}^{(m)}$ and diagonal covariance matrix $\Sigma_{i,t}$.  The $d^{th}$ diagonal element of $\Sigma_{i, t}$ is $c^2_{i,t} \sigma^2_{d}$, where $c_{i,t} = \alpha^{\frac{t-1+(i-1)T}{50T}}$.  The user must provide a choice of $\sigma_{d}$ for each dimension $\theta_{d}$ of $\theta$ as well as the parameter $\alpha$ that determines the cooling schedule for all dimensions together.  The formula for $c_{i, t}$ implies that after 50 IF2 iterations ($i = 51$, $t = 0$), the perturbation standard deviation for parameter $\theta_d$ will be $\alpha$ times the original perturbation standard deviation, $\sigma_{d}$.

The initial set of parameter particles is important for IF2's performance.  To generate these, the user must provide the mean, $\theta_{\mbox{init}}$, and standard deviations, i.e., a vector of square roots of diagonal elements of $\Sigma_{\mbox{init}}$. Note that the time step ``$T$'' for the initial particles $\theta_{0, T}^{(m)}$ is artificial so that generation of $\theta_{1, 0}^{(m)}$ from $h(\theta|\theta_{i-1, T}^{(m)})$ conforms to notation suitable for later iterations.

IF2 has been used for estimating parameters of stochastic differential equation models of disease dynamics (\citet{king15}, \citet{martinez15}, \citet{azman15}.)

\subsection{Particle MCMC methods}
\label{subsubsec:PMCMC}

Particle MCMC methods \citep[PMCMC; ][]{andrieu10} allow joint sampling from the posterior distribution of the states and the top-level parameters.  PMCMC takes advantage of the ability of certain particle filters to provide unbiased estimates of the (marginal) likelihood, that is, $\tilde{p}(y_{1:T}|\theta) \approx \int_{X}p(y_{1:T}|x_{1:T},\theta)p(x_{1:T}|\theta)dx_{1:T}$.   For example, the bootstrap filter and auxiliary filter \citep{pitt02} can both be used to provide unbiased estimates of the marginal likelihood, as detailed in Sections~\ref{subsubsec:Boot} and~\ref{subsubsec:Aux}.   Building upon the framework of \citet{andrieu09}, \citet{andrieu10} showed that using this approximation when calculating a Metropolis-Hastings acceptance ratio yields an ``exact approximate'' algorithm.  Although it seems to rely on an approximation, formally it samples from an auxiliary space of all states and indices sampled during the SMC run.  The marginal distribution of this enhanced space matches the desired target distribution, so it samples from exactly the correct target distribution based upon approximate likelihood calculations.  See \citet{dahlin19} for a gentle introduction.   Below, we detail the particle marginal Metropolis Hastings (PMMH) algorithm, one of three algorithms provided in \citet{andrieu10}.  The PMMH algorithm is available as a sampler within \pkg{nimble}'s general MCMC framework.

At each iteration $i$, the PMMH algorithm first proposes a value $\theta^{*}$ for the model parameters $\theta$ from a proposal distribution $q(\theta^{*}|\theta^{(i-1)})$.  Using this proposed value for $\theta$, a particle filter is then run, which provides an estimate $\tilde{p}(y_{1:T}|\theta^{*})$ of the likelihood given the proposed parameters.  This marginal likelihood estimate is used to calculate the Metropolis-Hastings acceptance probability for $\theta^{*}$.  By default, the algorithm saves the chain of $\theta$ samples.  If latent states are needed, an index $j$ is sampled from $\{1,\ldots,M\}$ using the final particle filter weights $\{\pi_{T}^{(m)}\}_{m=1}^{M}$ in PMMH iteration $i$, and the latent states $x_{1:T}^{(i)}$ for the $i^{th}$ PMMH iteration are set to the $j^{th}$ latent state sequence, $x_{1:T}^{(j)}$, from the $i^{th}$ run of the particle filter. Importantly, one does not need to record all states and indices used in the particle filter run, but one does need to retain $\tilde{p}(y_{1:T}|\theta^{*})$ if $\theta^{*}$ is accepted.
\begin{algorithm}[H]
	\caption{PMMH algorithm}
	\label{pmmh_algo}
	\begin{algorithmic}[1]
  \State Choose an initial value $\theta^{(0)}$ 
    \State Run a particle filter to obtain the marginal likelihood estimate $\tilde{p}^{(0)} = \tilde{p}(y_{1:T}|\theta^{(0)})$.
    \State After the last time step of the particle filter, draw $x_{1:T}^{(0)}\sim p(x_{1:T}|y_{1:T},\theta^{0})$ by sampling a particle with its full state history.
		\For{$i$ in $1:I$}
		\State Generate $\theta^{*} \sim q(\theta|\theta^{(i-1)})$ \label{line:pmmhResamp}
		\State Run a particle filter to obtain the marginal likelihood estimate $\tilde{p}(y_{1:T}|\theta^{*})$.
		\State After the last time step of the particle filter, draw $x_{1:T}^{*}\sim p(x_{1:T}|y_{1:T},\theta^{*})$ by sampling a particle with its full state history.
		\State Compute $a^{*}=1\wedge\frac{\tilde{p}(y_{1:T}|\theta^{*})p(\theta^{*})}{\tilde{p}^{(i-1)} p(\theta^{(i-1)})}\frac{q(\theta^{(i-1)}|\theta^{*})}{q(\theta^{*}|\theta^{(i-1)})}$\label{line:pmmhprob}
		\State Generate $r \sim unif(0,1)$
		\If{$a^{*}>r$}
			\State Set $\theta^{(i)}=\theta^{*}$,  $x_{1:T}^{(i)}=x_{1:T}^{*}$, and $\tilde{p}^{(i)} = \tilde{p}(y_{1:T}|\theta^{*})$
		\Else{}
			\State Set $\theta^{(i)}=\theta^{(i-1)}$, $x_{1:T}^{(i)}=x_{1:T}^{(i-1)}$, and $\tilde{p}^{(i)} = \tilde{p}^{(i-1)}$.
		\EndIf
		\EndFor
	\end{algorithmic}
\end{algorithm}

If the latent states, $x_{1:T}$, are not needed as part of the PMMH output, the steps to track and record them can simply be skipped.

PMCMC methods have been adopted in a wide variety of fields since their introduction.  Examples of research conducted using PMCMC include hydrology models \citep{vrugt13} and epidemiological models that incorporate genealogy \citep{rasmussen14}.

Like all MCMC algorithms, PMCMC methods can suffer from poor mixing.  In particular for PMCMC methods, such mixing problems can arise as a result of high variance of the likelihood estimates.  For example, an erroneously high likelihood estimate from the particle filter can make it difficult for subsequent proposals to be accepted even for nearby parameters.  These problems were studied theoretically by \citet{sherlock15} and \citet{doucet15} and are illustrated in the simulation exercise below.  Additionally, since high variances for the estimated likelihood occur especially for areas of low posterior density \citep{murray13}, it is important to provide reasonable initial values for the parameters, $\theta$.  One idea for obtaining reasonable initial values, similar to an idea from \citet{benton17}, is to use maximum a posteriori (MAP) estimates, if available, as starting values for PMCMC. 

\subsection{Ensemble Kalman filter}
\label{subsubsec:ENKF}

In Section~\ref{subsec:Filtering}, the Kalman filter was mentioned as providing an analytic solution to the filtering problem when working with a linear, Gaussian state-space model.  When using a model with non-linear transition equations or observation equations, however, the Kalman filter is no longer applicable.  One approximation to the filtering problem for Gaussian state-space models with non-linear transition or observation equations is the ensemble Kalman filter (EnKF), which uses a particle representation of the latent states at each time point.  

Although the EnKF uses particles to represent filtering distributions, as do the particle filters described in Sections~\ref{subsubsec:Boot} and ~\ref{subsubsec:Aux},  it updates the latent state particles using a fundamentally different approach than those methods.  Whereas bootstrap and auxiliary particle filters update particles via a re-weighting and re-sampling framework,  the EnKF first propagates particles forward using the transition equation, and then shifts particles towards the filtering distribution using an approximation to the Kalman gain matrix from the original Kalman Filter \citep{mandel09}. 

The EnKF assumes the following forms for the transition and observation equations:
\begin{align}
x_{t} &= M(x_{t-1})+w_{t} \label{eq:1}\\ 
y_{t} &= D(x_{t})+v_{t} \label{eq:2}
\end{align}
where $w_{t}$ and $v_{t}$ are independent, normally distributed error terms with covariance matrices $Q_{t}$ and $R_{t}$ respectively.  At time $t-1$, assume that we have a sample $\{x_{t-1}^{(m)}\}_{m=1}^M$ of  particles from $p(x_{t-1}|y_{1:t-1})$.  Each particle is propagated forward according to Equation~\ref{eq:1}, with random draws for $w_t$, giving $\tilde{x}_{t}^{(m)}$.  In addition, the mean observation from each $\tilde{x}_{t}^{(m)}$ is calculated as $\tilde{y}_{t}^{(m)} = D(\tilde{x}_{t}^{(m)})$.   The idea behind the EnKF is to use these samples to approximate the covariance between latent states and observations, as well as the covariance among observation dimensions, at time $t$.  From these, one can approximate the Kalman gain matrix, $\tilde{G}_{t}$, and apply a simulated version of the Kalman update step as follows: \[x_{t}^{(m)} = \tilde{x}_{t}^{(m)} + \tilde{G}_{t}(y_{t} + v^{(m)}_{t} - \tilde{y}_{t}^{(m)}),\] where $v_{t}^{(m)} \sim N(0, R_{t})$ are simulated observation errors.  The $x_{t}^{(m)}$ values form a sample of the approximated filtering distribution at time $t$, i.e., $p(x_{t}|y_{1:t})$.

The covariance between latent states and observations is calculated as the covariance between $e_{t}^{x}=(\tilde{x}_{t}^{(1)} - \bar{\tilde{x}}_{t},\ldots,\tilde{x}_{t}^{(M)} - \bar{\tilde{x}}_{t})$ and $e_{t}^{y}=(\tilde{y}_{t}^{(1)} - \bar{\tilde{y}}_{t},\ldots,\tilde{y}_{t}^{(M)} - \bar{\tilde{y}}_{t}))$.  This, along with a similarly approximated covariance of observations, is used to construct the approximate Kalman gain matrix $\tilde{G}_{t}$.  A more detailed overview of the EnKF, and how it relates to the Kalman Filter, can be found in \citet{gillijns06} and \citet{katzfuss16}.

\begin{algorithm}[H]
  \caption{Ensemble Kalman filter}
	\label{enkf_algo}
  \begin{algorithmic}[1]
    \For{$m$ in $1:M$}
     \State Generate $x_{0}^{(m)}\sim f(x_{0})$
    \EndFor 
		\For{$t$ in $1:T$}
		\For{$m$ in $1:M$}
		\State Generate $\tilde{x}_{t}^{(m)}\sim f(x_{t}|x_{t-1}^{(m)})$
    \State Calculate $\tilde{y}_{t}^{(m)} = D(\tilde{x}_{t}^{(m)})$
    \EndFor 
    \State Calculate $e_{t}^{x}=(\tilde{x}_{t}^{(1)} - \bar{\tilde{x}}_{t},\ldots,\tilde{x}_{t}^{(M)} - \bar{\tilde{x}}_{t})$
    \State Calculate $e_{t}^{y}=(\tilde{y}_{t}^{(1)} - \bar{\tilde{y}}_{t},\ldots,\tilde{y}_{t}^{(M)} - \bar{\tilde{y}}_{t})$
    \State Calculate $\tilde{P}_{t}^{xy} = \frac{1}{M-1}e_{t}^{x}(e_{t}^{y})^{'}$
    \State Calculate $\tilde{P}_{t}^{yy} = \frac{1}{M-1}e_{t}^{y}(e_{t}^{y})^{'} + R_{t}$
    \State Calculate $\tilde{G}_{t} = \tilde{P}_{t}^{xy}(\tilde{P}_{t}^{yy})^{-1}$
     \For{$m$ in $1:M$}
      \State Generate $v_{t}^{(m)} \sim N(0, R_{t})$
      \State Calculate $x_{t}^{(m)} = \tilde{x}_{t}^{(m)} + \tilde{G}_{t}(y_{t}+v_{t}^{(m)}-\tilde{y}_{t}^{(m)})$
      \EndFor
	\EndFor
	\end{algorithmic}
\end{algorithm}

Note that the multiplication of $e_{t}$ terms in steps 11 and 12 are matrix multiplications.

We remark that although the EnKF assumes normally distributed error terms for Equations~\ref{eq:1} and \ref{eq:2}, the filter has been shown to be robust to each of those assumptions \citep{katzfuss16}.
Also, although newer variations of the EnKF exist that may perform better for some models, such as the Ensemble Adjusted Kalman Filter \citep{anderson01}, we have provided only a basic version the EnKF algorithm in \pkg{nimble}.  Users are welcome to modify this basic EnKF if a different version is desired - see Section~\ref{sec:nimbleFuncs} for an overview of creating and modifying particle filters in \pkg{nimble}.  Recently, the EnKF has been used extensively for atmospheric data modeling \citep{houtekamer16}, as well as for problems in petroleum modeling \citep{heidari13} and geophysics \citep{bocher18}.

\section{Creating and manipulating models in nimble}\label{subsec:nimbleModels}

The workflow of modeling and conducting inference in \pkg{nimble} is somewhat unique as compared to other statistical modeling software.  Thus, before demonstrating \pkg{nimble}'s SMC features, we will first show in this section how to create and work with \pkg{nimble}'s model objects.  Section~\ref{subsec:fixedParam} describes \pkg{nimble}'s SMC methods for latent state inference in state-space models. Section~\ref{subsec:unknownParam} describes \pkg{nimble}'s SMC methods for top-level parameter inference.   A supplement to the paper includes a full \proglang{R} script of all code shown below.

The \pkg{nimble} package uses the \proglang{BUGS} language to specify hierarchical statistical models.  We will not describe the \proglang{BUGS} language here -- interested readers can find a brief overview in the \pkg{nimble} User Manual \citep{nimble-manual}, or a more detailed guide in \citet{lunn12}.  
Unlike other packages that use dialects of the \proglang{BUGS} language (\pkg{WinBUGS} and \pkg{JAGS}), \pkg{nimble} creates model objects, via the \code{nimbleModel} function, which can be queried and manipulated by the user.  To introduce \pkg{nimble}'s modeling framework, we will use a linear Gaussian state-space model in which all parameters are fixed.  

Let $y_{t}$ be the observed data at time $t$, let $x_{t}$ be the latent state at time $t$, and suppose we have 10 times.  The model is:

\begin{align*}
x_{0} &\sim N(0, 1)\\
x_{t} &\sim N(0.8x_{t-1}, 1) \,\,\,\, \text{for}\,\,\,\,\, t=1,\ldots,10  \\
y_{t} &\sim N(x_{t}, 0.5) \,\,\,\, \text{for}\,\,\,\,\, t=1,\ldots,10  \\
\end{align*}

where $N(\mu,\sigma^{2})$ denotes the Normal distribution with mean $\mu$ and variance $\sigma^{2}$.  Although this example model is relatively simple, it will serve to demonstrate the process of building and working with \proglang{BUGS} models in \pkg{nimble}.

 Code written in the \proglang{BUGS} language is entered via the \code{nimbleCode} function.  For example, \proglang{BUGS} code for the linear Gaussian model is entered like this: 

\begin{Schunk}
\begin{Sinput}
R> library("nimble")
R> exampleCode <- nimbleCode({
+     x0 ~ dnorm(0, var = 1)
+     x[1] ~ dnorm(.8 * x0, var = 1)
+     y[1] ~ dnorm(x[1], var = .5)
+     for(t in 2:10){
+       x[t] ~ dnorm(.8 * x[t-1], var = 1)
+       y[t] ~ dnorm(x[t], var = .5)
+     }
+   })
\end{Sinput}
\end{Schunk}
 Each line of \proglang{BUGS} code declares one or more nodes in the model.  For example, this model has the nodes \code{x0}, \code{x[1]}, $\ldots$, \code{x[10]}, \code{y[1]}, $\ldots$, \code{y[10]}.
 
 Once the model code has been written, a model object in \proglang{R} can be created using the \code{nimbleModel} function.   Data and initial values can optionally be provided at this step.  For example:
\begin{Schunk}
\begin{Sinput}
R> simulatedData <- c(-0.9,  1.6,  0.6,  1.3,  1.5, 0.3, -0.8, -1.3,  0.5,
+   1.1)
R> exampleModel <- nimbleModel(code = exampleCode,
+   data = list(y = simulatedData),
+   inits = list(x0 = 0))
\end{Sinput}
\end{Schunk}

The  \code{exampleModel} object is an \proglang{R} reference class object.  \code{exampleModel} has a field for each of the variables in our \proglang{BUGS} model.  For example, we can look at the value of the \code{y[3]} node by calling
\begin{Schunk}
\begin{Sinput}
R> exampleModel$y[3]
\end{Sinput}
\begin{Soutput}
[1] 0.6
\end{Soutput}
\end{Schunk}
 
 Additionally,  \code{exampleModel} provides \code{simulate} and \code{calculate} methods.  The \code{simulate} method takes a vector of node names as an argument, simulates values for these nodes conditioned upon the values of other nodes in the model that are parents of the nodes being simulated, and  stores the simulated values in the corresponding model variables.  The following code simulates values for \code{x0}, \code{x[1]}, $\ldots$, \code{x[10]}.  We note that \code{x[1]} is simulated conditional upon the current value of \code{x0}, \code{x[2]} is simulated conditional upon the simulated value of \code{x[1]}, and so on.  Note that the call to the \code{getNodeNames} method, using \code{includeData = FALSE}, will return the names of all non-data nodes in our model in an order that is valid for calculation or simulation.
   
\begin{Schunk}
\begin{Sinput}
R>  exampleModel$simulate(nodes =
+  exampleModel$getNodeNames(includeData = FALSE))
R>  exampleModel$x
\end{Sinput}
\begin{Soutput}
 [1] -0.31751972 -1.08964439  0.72356529  0.90836000 -0.09378038
 [6]  0.41240475  1.06824850  1.43038015  0.83891574  2.18291376
\end{Soutput}
\end{Schunk}
 
 The \code{calculate} method takes a vector of node names as an argument, and returns the summed log probability of all of the given stochastic nodes.  The following line demonstrates calculating $\sum_{t=1}^{10}log(g(y_{t}|x_{t}))$. 
\begin{Schunk}
\begin{Sinput}
R> exampleModel$calculate(nodes = 'y')
\end{Sinput}
\begin{Soutput}
[1] -28.25143
\end{Soutput}
\end{Schunk}
 \pkg{nimble} model objects contain additional methods that can provide details about the model and allow for more fine-grained manipulation.  These are described in Chapter 6 of the \pkg{nimble} User Manual \citep{nimble-manual}.
 
 Although models and algorithms in \pkg{nimble} can run entirely within \proglang{R}, they are generally compiled into \proglang{C++} for greatly increased efficiency.  Compilation is achieved through the \code{compileNimble} function, which takes as arguments one or more model objects (or objects created by calls to \code{nimbleFunction}, described in Section~\ref{sec:nimbleFuncs}).  \code{compileNimble} generates model-specific \proglang{C++}, compiles and loads the result, and returns a version of the model object with the same fields and methods, but where the methods are now executed in compiled \proglang{C++} instead of \proglang{R}.
 
\begin{Schunk}
\begin{Sinput}
R>   C_exampleModel <- compileNimble(exampleModel)
\end{Sinput}
\end{Schunk}
 
 \code{compileNimble} can also be used to compile algorithms in \pkg{nimble}, which will use compiled versions of the model.
  
  Generally, analysis in \pkg{nimble} will start with writing model code and creating a model object.  If available, initial values and data values should be provided to the model.  A user then has a choice of inference methods: in addition to the SMC methods documented in this paper, \pkg{nimble} has a flexible MCMC system for posterior inference, an MCEM algorithm, and cross-validation methods.  A user may also wish to write their own method using \pkg{nimble}'s programming system.  Once an algorithm has been chosen, the user can compile both the model and the algorithm to \proglang{C++} using the \code{compileNimble} function.  Finally, the user can run the algorithm and analyze the results.  

  In Section~\ref{subsec:fixedParam}, we introduce \pkg{nimble}'s bootstrap filter, auxiliary filter, and ensemble Kalman filter by applying them to the state-space model created above.  

 \section{Filtering given fixed parameters}\label{subsec:fixedParam}
 Now that we have built a \code{model} object for the example model, we can use algorithms from \pkg{nimble}'s library.   These algorithms are all written as \code{nimbleFunctions} using \pkg{nimble}'s programming system embedded in R.     We begin by demonstrating the use of the bootstrap filter (Section~\ref{subsubsec:Boot}) to estimate the filtering distribution $p(x_t|y_{1:t})$. 
\begin{Schunk}
\begin{Sinput}
R> exampleBootstrapFilter <- buildBootstrapFilter(exampleModel, nodes = 'x',
+   control = list(saveAll = TRUE, thresh = .9))
\end{Sinput}
\end{Schunk}
   The \code{buildBootstrapFilter} function builds a bootstrap filter for the model given in the first argument.  The \code{nodes} argument gives the name (or names) of the latent states to be filtered.   Importantly, the latent states must have the same dimension at each time point.  The algorithm parameters, packaged in the control list, include \code{saveAll} (should filtered state estimates be saved from all time points, or from just the last one) and \code{thresh} (a threshold for resampling, labeled $\tau$ in Section~\ref{subsubsec:Aux}).  Additional arguments to the control list can be found by calling \code{help(buildBootstrapFilter)}.  One control list parameter of note is \code{smoothing}, which defaults to \code{FALSE}.  If \code{smoothing = TRUE}, the particles returned from the algorithm will be samples from the smoothing distribution $p(x_{1:T}|y_{1:T})$.  Along with similar functions that appear below, \code{buildBootstrapFilter} is actually a \code{nimbleFunction}, meaning it is written in \pkg{nimble}'s algorithm programming system.
   
     The bootstrap filter in \pkg{nimble} sets the proposal distribution $ q(x_{t}|x_{t-1}^{(m)},y_{t})$ to be the transition equation  $f({x}_{t}|{x}_{t-1}, \theta)$.  A user wishing to use a different proposal distribution would need to copy and modify \code{buildBootstrapFilter} (see Section~\ref{sec:nimbleFuncs}).
  
   After the bootstrap filter has been built, it can be run in \proglang{R} by calling the \code{run} method of the filter, taking the number of particles to use as an argument, and returning an estimate of the log likelihood of the data.  
\begin{Schunk}
\begin{Sinput}
R> exampleBootstrapFilter$run(100)
\end{Sinput}
\begin{Soutput}
[1] -15.16996
\end{Soutput}
\end{Schunk}
Running an algorithm uncompiled allows for easy testing and debugging of algorithm logic using, for example, \proglang{R}'s \code{browser()} and \code{trace()} functions.  Once an algorithm has been successfully constructed in \proglang{R}, it can be compiled into \proglang{C++} for efficient execution.  Below, we compile the bootstrap filter algorithm using the \code{compileNimble} function, as seen in Section~\ref{subsec:nimbleModels}, and run the compiled filter using 10,000 particles. Note that the model must be compiled before or in the same step as the algorithm.  The \code{exampleModel} here was compiled in Section~\ref{subsec:nimbleModels}.
\begin{Schunk}
\begin{Sinput}
R> CexampleBootstrapFilter <- compileNimble(exampleBootstrapFilter,
+  project = exampleModel)
R> CexampleBootstrapFilter$run(10000)
R> bootstrapFilterSamples <- as.matrix(CexampleBootstrapFilter$mvEWSamples)
\end{Sinput}
\end{Schunk}
 The bootstrap filter, like most particle filters in \pkg{nimble}, saves two arrays with samples from the filtering distribution.  One array, named \code{mvEWSamples}, contains equally weighted samples from the filtering distribution.  The second array, \code{mvWSamples}, contains non-equally weighted samples from the filtering distribution along with log weights for each sample.  These arrays can be easily converted to \proglang{R} matrices via the \code{as.matrix} function as shown above.
 
 Next, we demonstrate \pkg{nimble}'s auxiliary particle filter algorithm (Section~\ref{subsubsec:Aux}). The auxiliary particle filter is constructed by the \code{buildAuxiliaryFilter} function.    Users can choose between two lookahead functions: one that uses a simulation from the transition equation $\tilde{x}^{(m)}_{t|t-1}\sim f(x_{t}|x_{t-1}^{(m)})$, and one that uses the expected value of the transition equation $\tilde{x}^{(m)}_{t|t-1} = E(x_{t}|x_{t-1}^{(m)})$, via the \code{lookahead} control list argument.  The expected value lookahead function can only be used for state-space models with normal or multivariate normal transition equations.
 
\begin{Schunk}
\begin{Sinput}
R> exampleAuxiliaryFilter <- buildAuxiliaryFilter(exampleModel, nodes = 'x',
+  control = list(saveAll = TRUE, lookahead = 'mean'))
R> CexampleAuxiliaryFilter <- compileNimble(exampleAuxiliaryFilter,
+  project = exampleModel, resetFunctions = TRUE)
R> CexampleAuxiliaryFilter$run(10000)
R> auxiliaryFilterSamples <- as.matrix(CexampleAuxiliaryFilter$mvEWSamples)
\end{Sinput}
\end{Schunk}
 
The \code{resetFunctions = TRUE} option is helpful when compiling a second algorithm for a model.  See the \pkg{nimble} User Manual for more details.

  The final method we demonstrate for models with fixed parameters is the ensemble Kalman filter, which can be built via a call to \code{buildEnsembleKF}.  Note that the ensemble Kalman filter, as described in Section~\ref{subsubsec:ENKF}, does not produce weights with its particle estimates.  Thus there is only one output array, named \code{mvSamples}. Additionally, we note that the ensemble Kalman filter in \pkg{nimble} will only work with Gaussian noise in the process and observations, although the mean state dynamics or observations can be non-linear. 
\begin{Schunk}
\begin{Sinput}
R> exampleEnsembleKF <- buildEnsembleKF(exampleModel, nodes = 'x',
+ control = list(saveAll = TRUE))
R> CexampleEnsembleKF <- compileNimble(exampleEnsembleKF,
+ project = exampleModel, resetFunctions = TRUE)
R> CexampleEnsembleKF$run(10000)
R> EnKFSamples <- as.matrix(CexampleEnsembleKF$mvSamples)
\end{Sinput}
\end{Schunk}
 Since our example model has normal transition and observation equations, the filtering distribution can also be calculated exactly using the Kalman filter  \citep{kalman60}.  Below, we use the \pkg{dlm} package \citep{petris10} to apply a Kalman filter to our model and compare the exact filtering distribution provided by the Kalman filter to the approximate filtering distributions given by the bootstrap filter, auxiliary particle filter, and EnKF.  Note that the quantiles in Figure~\ref{fig:exampleDLM_Chunk} align almost exactly for all filters. 

\begin{Schunk}
\begin{figure}[H]

{\centering \includegraphics[width=.69\linewidth]{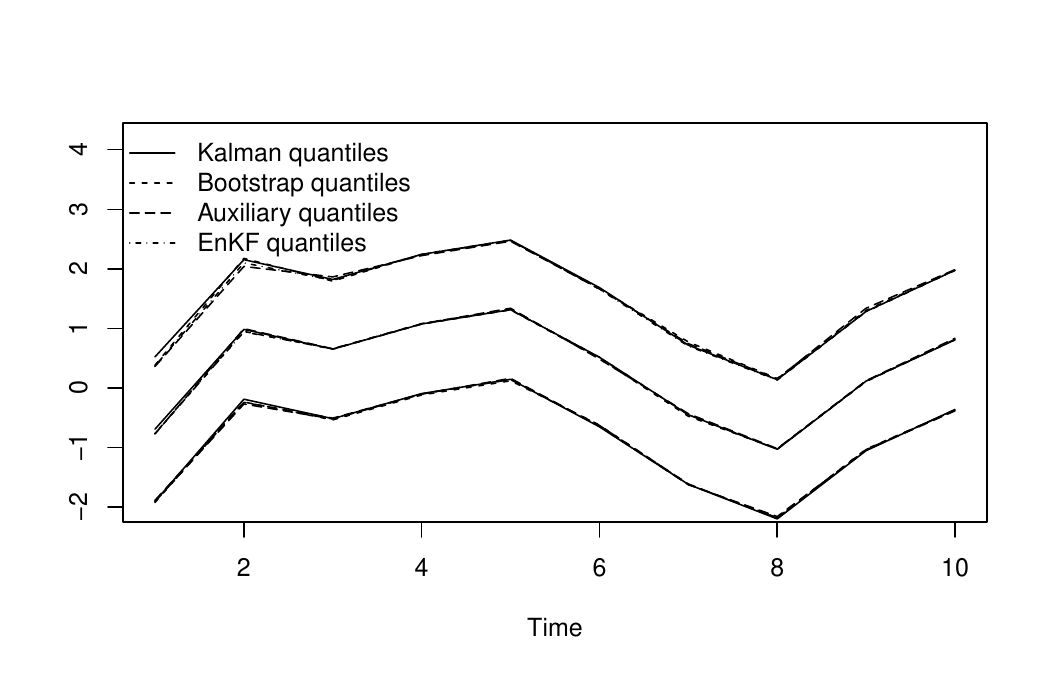} 

}

\caption[2.5\%, 50\% and 97.5\% quantiles of the filtering distribution for the Kalman filter and nimble's particle filters]{2.5\%, 50\% and 97.5\% quantiles of the filtering distribution for the Kalman filter and nimble's particle filters.}\label{fig:exampleDLM_Chunk}
\end{figure}
\end{Schunk}
 
 \section{Inference on models with unknown parameters}\label{subsec:unknownParam}
 
The example model in the previous section had no unknown parameters -- an uncommon scenario with real data.  We next demonstrate \pkg{nimble}'s  PMCMC and IF2 algorithms, which can be used to estimate the unknown top-level parameters in a Bayesian and frequentist framework, respectively.  For PMCMC, we use a stochastic volatility example following package \pkg{stochvol}.  For IF2, we use a linear random walk model of Nile river flows with a single changepoint, following the use of this example by \citet{durbin12} and in packages \pkg{dlm} and \pkg{FKF}.  For both methods, we present some computational experiments illustrating the roles of some tuning parameters: proposal scale for PMCMC, cooling parameter for IF2, and number of particles for both.  While IF2 can handle nonlinear and/or non-Gaussian models, use of the linear, Gaussian example facilitates our computational experiment by allowing easy calculation of the correct likelihood.  Both examples are also chosen to facilitate verification of correct results, by comparison to \pkg{stochvol}'s own MCMC for our PMCMC example, and by exact maximum likelihood estimation for our IF2 example.

 \subsection{Particle Marginal Metropolis-Hastings}\label{subsubsec:PMMH}

 \pkg{nimble}'s PMCMC uses the particle marginal Metropolis-Hastings (PMMH) method.  This can take advantage of \pkg{nimble}'s existing MCMC framework, in which the user can assign different samplers to different nodes in a model.    For a full description of \pkg{nimble}'s MCMC design, see Chapter 7 of the \pkg{nimble} User Manual \citep{nimble-manual}.  The PMMH sampler in \pkg{nimble} uses a normal proposal distribution, and can sample either scalar parameters (using the \code{RW\_PF} sampler) or vectors of parameters (using the \code{RW\_PF\_block} sampler). Multiple such samplers can also be combined.  Before setting up PMMH, we introduce the stochastic volatility example.
 
\subsubsection{Stochastic volatility example}\label{subsubsec:stochVol}
 Stochastic volatility models are widely used for time-series of the log returns of financial assets such as stocks.  The main idea is to model volatility (standard deviation of log returns) as an unobserved autoregressive process.  Our example is drawn from the \proglang{R} package \pkg{stochvol} \citep{kastner16, kastner19}, where it is described in detail in the vignette ``Dealing with Stochastic Volatility in Time Series Using the R Package stochvol''.   \citet{pitt99} use a related auxiliary particle filter example.  We reparameterize relative to the  \pkg{stochvol} example as noted below.

Let $r_{t}$ be the exchange rate at time $t$, and define $y_{t}$ as the daily log return, that is, $y_{t} = (log(r_{t}) - log(r_{t-1}))$ for $t=2,\ldots,T$.  The stochastic volatility model is:
\begin{align*}
   x_{t} &\sim N(\phi x_{t-1}, \sigma^{2}).\\
   y_{t} &\sim N(0, (\beta \exp(x_{t} / 2))^2)\\
\end{align*}

In this model, $\beta$ can be interpreted as the constant volatility, while $x_{t}$ is the latent, evolving log of volatility squared. For the distribution of the initial state, we use the stationary distribution:
\begin{align*}
x_1 &\sim N(0, \sigma^{2} / (1-\phi^2)).
\end{align*}

Prior distributions are placed on the parameters $\beta$, $\phi$, and $\sigma$ as follows, using the same choices as the \pkg{stochvol} vignette, specifically:
\begin{align*}
 \phi^{*} &\sim B(20, 1.1),\\
 \phi &= 2 \phi^{*} - 1,\\
 \Omega &\sim  \frac{r^s  e^{s \Omega} e^{-r e^\Omega}}{ \Gamma(s) }, s = 0.5, r = 5,\\
 \sigma^2 &= \exp(\Omega),\\
 \mu &\sim  N(-10, 1),\\
 \beta &= \exp(\mu / 2),
 \end{align*}

where the density function given for $\Omega$ is such that $\exp(\Omega)$ follows a gamma distribution with shape $s$ and rate $r$.  

A normal prior is placed on $\mu = 2 \log \beta$.  This is moderately informative, based on knowledge of typical mean volatility.   The choice of $-10$ for the mean would be different if we used volatility on a percent scale ($100 \times y_t$ instead of $y_t$).  An informative beta prior is placed on $\phi^{*} = (\phi  + 1)/2$.  According to \pkg{stochvol}, informative priors are common because the data are only weakly informative for $\phi$, yet there is domain knowledge of typical values.  Finally, a moderately-informative gamma prior is placed on $\sigma^2$, as done in \pkg{stochvol}, but we parameterize that prior in terms of $\Omega = \log \sigma^2$.  The latter choice is made because it facilitates better mixing when block-sampling all three parameters.  In order to define $\Omega$ to follow a distribution corresponding to a gamma on $\sigma^2$, we use \pkg{nimble}'s extensibility for writing new distributions.  While these choices for priors are somewhat arbitrary, sticking to the choices of the \pkg{stochvol} vignette while reparameterizing provides an illustration of the usefulness of \pkg{nimble}'s extensibility.

Code for the stochastic volatility model is:
 
\begin{Schunk}
\begin{Sinput}
R>  stochVCode <- nimbleCode({
+   x[1] ~ dnorm(0, sd = sigma / sqrt(1-phi*phi))
+   y[1] ~ dnorm(0, sd = beta * exp(0.5 * x[1]))
+   for(t in 2:T){
+     x[t] ~ dnorm(phi * x[t-1], sd = sigma)
+     y[t] ~ dnorm(0, sd = beta * exp(0.5 * x[t]))
+   }
+   phi <- 2 * phiStar - 1
+   phiStar ~ dbeta(20, 1.1)
+   logsigma2 ~ dgammalog(shape = 0.5, rate = 1/(2*0.1)) ## This is Omega
+   sigma <- exp(0.5*logsigma2)
+   mu ~ dnorm(-10, sd = 1) ## It matters whether data are converted to % or not.
+   beta <- exp(0.5*mu)
+ })
\end{Sinput}
\end{Schunk}

Note that \code{dgammalog} is the opposite of a
log-gamma distribution.  That is, $\exp(\Omega)$ (rather than $\log(\Omega)$) follows a
gamma distribution. Code for the density and random simulation functions
for \code{dgammalog} is:
\begin{Schunk}
\begin{Sinput}
R> dgammalog <- nimbleFunction(
+   run = function(x = double(), shape = double(),
+                  rate = double(),log = integer(0, default = 0)) {
+     logProb <- shape * log(rate) + shape * x - rate * exp(x) - lgamma(shape)
+     if(log) return(logProb)
+     else return(exp(logProb))
+     returnType(double())
+   }
+ )
R> 
R> rgammalog <- nimbleFunction(
+   run = function(n = integer(),
+                  shape = double(), rate = double()) {
+     xg <- rgamma(1, shape = shape, rate = rate)
+     return(log(xg))
+     returnType(double())
+   }
+ )
\end{Sinput}
\end{Schunk}

We use as data exchange rates for the Euro (EUR) quoted in U.S. Dollars (USD) starting after January 1st, 2010, and continuing until the end of the time-series, 582 days after that.  This data set can be found in \pkg{stochvol} \citep{kastner16}, along with the function \code{logret} to calculate log returns.

\begin{Schunk}
\begin{Sinput}
R> library("stochvol")
R> data("exrates")
R> y <- logret(exrates$USD[exrates$date > '2010-01-01'], demean = TRUE)
R> T <- length(y)
\end{Sinput}
\end{Schunk}

We next create and compile a \code{nimbleModel} object for the above \proglang{BUGS} code, using as starting values $\mu = -10$, $\phi^{*} = .99$, and $\log \sigma^2 = -5.52$, and providing $T$ as a constant.
\begin{Schunk}
\begin{Sinput}
R> stochVolModel <- nimbleModel(code = stochVCode,
+  constants = list(T = T), data = list(y = y),
+  inits = list(mu = -10, phiStar = .99, logsigma2 = log(.004)))
R> CstochVolModel <- compileNimble(stochVolModel)
\end{Sinput}
\end{Schunk}

\subsubsection{Building and running PMMH}\label{subsubsec:build-run-pmmh}

To implement the PMMH algorithm, we first set up an MCMC configuration for our stochastic volatility model using the \code{configureMCMC} function.  We use the \code{monitors} argument to set the list of nodes for which we want \pkg{nimble} to return posterior samples. 

The PMMH sampler can be added to the MCMC configuration with a call to the \code{addSampler} method.  Additional options to customize the sampler can be specified within the \code{control} list.
\begin{Schunk}
\begin{Sinput}
R> stochVolMCMCConf <- configureMCMC(stochVolModel, nodes = NULL,
+   monitors = c('mu', 'beta', 'phiStar', 'phi', 'logsigma2', 'sigma'))
R> auxpf <- buildAuxiliaryFilter(stochVolModel, 'x', 
+   control = list(saveAll = FALSE, smoothing = FALSE, initModel = FALSE))
R> h <- 1
R> propSD <- h * c(0.089, 0.039, 1.45)
R> m <- 100
R> stochVolMCMCConf$addSampler(target = c('mu', 'phiStar', 'logsigma2'),
+   type = 'RW_PF_block', control = list(propCov = diag(propSD^2),
+                                        pf = auxpf,
+                                        adaptive = FALSE,
+                                        pfNparticles = m,
+                                        latents = 'x'))
\end{Sinput}
\end{Schunk}

In the last step above, we add a block random walk sampler with a
multivariate normal proposal distribution by specifying \code{type =
  `RW\_PF\_block'}.  This sampler is used to obtain posterior samples of
the \code{target} parameters: \code{mu}, \code{phiStar}, and
\code{logsigma2}.  Although these are the parameters with priors, we give results below for the transformed parameters of interest, namely $\mu$, $\phi$, and $\sigma$.  In this example, we chose to build the auxiliary particle filter first and provide it as a \code{control} argument to the sampler.  Doing it this way can allow multiple samplers to share the same particle filter.  If there is only one sampler (as here), one can alternatively use \code{pf = `auxiliary'} to have the sampler itself create the particle filter.

Once the PMMH sampler is added to the MCMC configuration, the algorithm can be built using the \code{buildMCMC} function and then compiled.   Posterior samples are stored in \code{cMCMC\$mvSamples}.  Below we demonstrate running \pkg{nimble}'s PMMH algorithm for 50,000 iterations.  
  
\begin{Schunk}
\begin{Sinput}
R> stochVolMCMC <- buildMCMC(stochVolMCMCConf)
R> cMCMC <- compileNimble(stochVolMCMC, project = stochVolModel,
+   resetFunctions = TRUE)     
R> cMCMC$run(50000)
R> samples <- as.matrix(cMCMC$mvSamples)
\end{Sinput}
\end{Schunk}

\begin{figure}
  \includegraphics[width=6in]{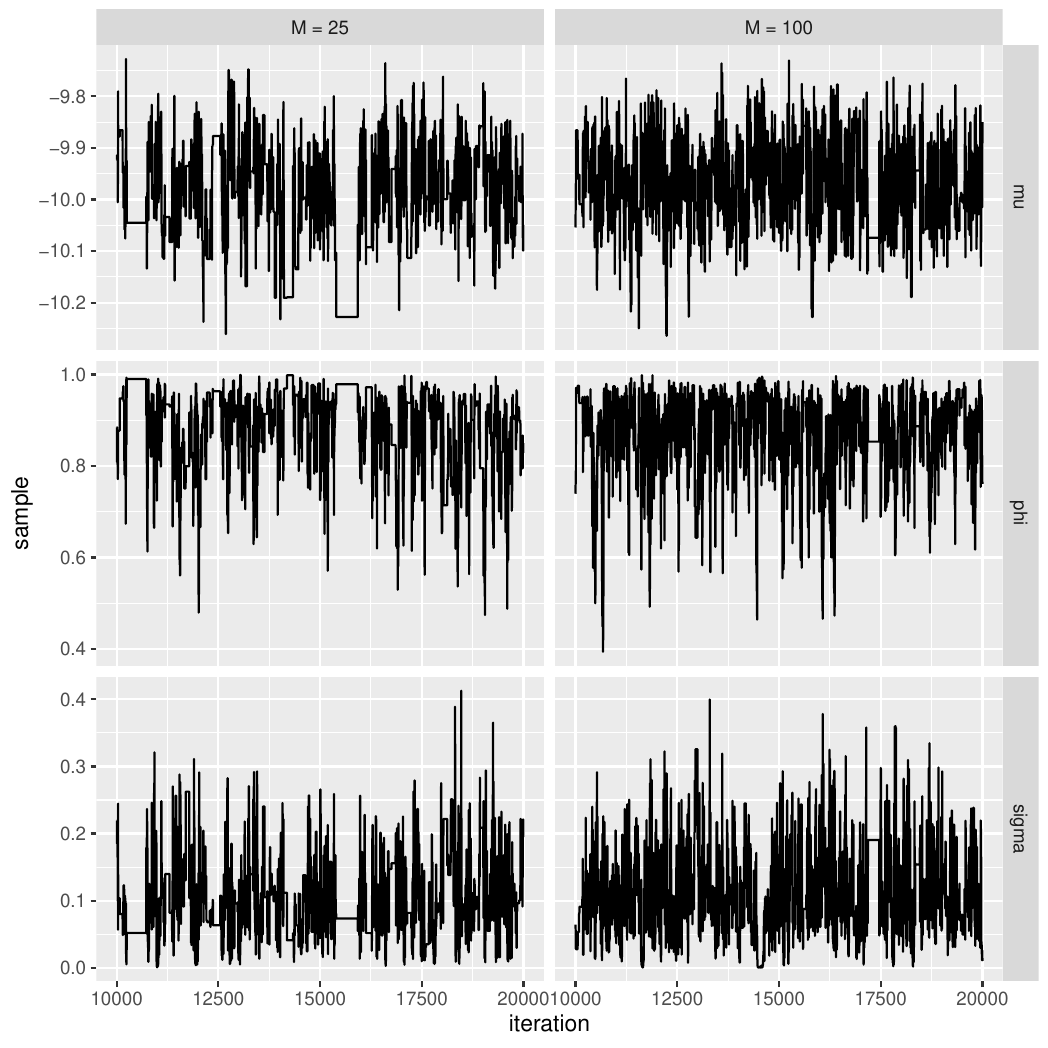}
  \caption{Traceplots of PMCMC with M=25 (left) or 100 (right) particles in
  each run of the particle filter.  10000 of 50000 iterations are shown.}
  \label{fig:PMCMCtraceplot}
\end{figure}

Figure \ref{fig:PMCMCtraceplot} shows traceplots of 10000 iterations from this MCMC (right panel, M = 100).  For comparisons below, traceplots with fewer particles (left panel, M = 25) are shown.  These display more pronounced ``stickiness'', where the MCMC gets stuck for many subsequent iterations because the likelihood was overestimated when accepting a proposal in a given iteration, as discussed in Section~\ref{subsec:tune-pmcmc}  In harder applications, larger numbers of particles will typically be needed.

As mentioned in Section~\ref{subsec:fixedParam}, \pkg{nimble} offers two choices for the lookahead function of the auxiliary particle filter, which can be specified for use in a PMMH sampler via the \code{pfLookahead} control list argument or specified when building the particle filter directly.  However, there may be cases where another choice of lookahead function is desired, or even where a user would like to use a filtering algorithm other than the bootstrap and auxiliary filters.  To accommodate these scenarios, \pkg{nimble} allows for user-defined filtering algorithms within its PMCMC sampler.  Thus a user could, for example, modify \pkg{nimble}'s auxiliary particle filter code to have a customized lookahead function and use this modified sampler to produce likelihood estimates in PMCMC.  More information on creating user-defined filtering algorithms can be found in Section~\ref{sec:nimbleFuncs}, and information on using such algorithms in PMCMC can be found in Section 8.1.2  of \pkg{nimble}'s User Manual \citep{nimble-manual}.  \citet{pitt99} show that a filter that uses a good look-ahead method will tend to have a lower mean squared error than either the bootstrap or other auxiliary particle filters.  Such an adapted filter would be a prime candidate for a user-defined filter in \pkg{nimble}'s particle MCMC samplers.
  
Similarly, \pkg{nimble}'s \code{RW\_PF} and \code{RW\_PF\_block} samplers use normal proposal distributions, but \pkg{nimble} provides a a straightforward system for writing new samplers.  See Section 13.5 of the User Manual for information on user-defined MCMC samplers.

\subsection{Tuning PMCMC}\label{subsec:tune-pmcmc}
 
Next we illustrate tuning considerations for PMMH, which has aspects distinct from other MCMC methods.  An intuitive interpretation of PMMH is that it uses SMC to approximate the likelihood in the Metropolis-Hastings acceptance probability.  However, its theoretical justification is actually different.  PMMH is valid because the entire set of simulated states and indices (during resampling) represent auxiliary variables in a complicated enhanced space for the MCMC.  The marginal distribution of the variables of interest matches the target distribution, which justifies sampling in the enhanced space.  This is discussed in \citet{andrieu09}, \citet{andrieu10}, \citet{doucet15}, \citet{sherlock15}, and \citet{dahlin19}.

This insight leads to what might be surprising performance considerations.  From the intuitive interpretation, one might be tempted to increase the number of SMC particles in order to make the approximation in the acceptance probability more accurate.  However, the justification of the method is valid even for small numbers of SMC particles, leading to the label ``exact approximation'' \citep{andrieu10}.  The risk with a small number of particles is not that the algorithm will be formally invalid but rather that it will get stuck due to high variance in the SMC likelihood approximation.  If a single run of the SMC yields an extremely large likelihood by chance, it will typically be accepted, after which many proposals may be rejected because the acceptance ratio will be low.  Increasing SMC sample size decreases the variance of likelihood approximations, which reduces the chance of getting stuck.  However, computational cost scales approximately linearly with SMC sample size, so it may be better to use fewer SMC samples and be able to run more iterations. Theory about these considerations is developed by \citet{sherlock15} and \citet{doucet15}.   In summary, whereas sample size is chosen for accuracy of the likelihood and state approximations in vanilla SMC, it is chosen to balance mixing and computational cost when the SMC is used in PMMH.

\begin{figure}
\centering
  \includegraphics[width=6in]{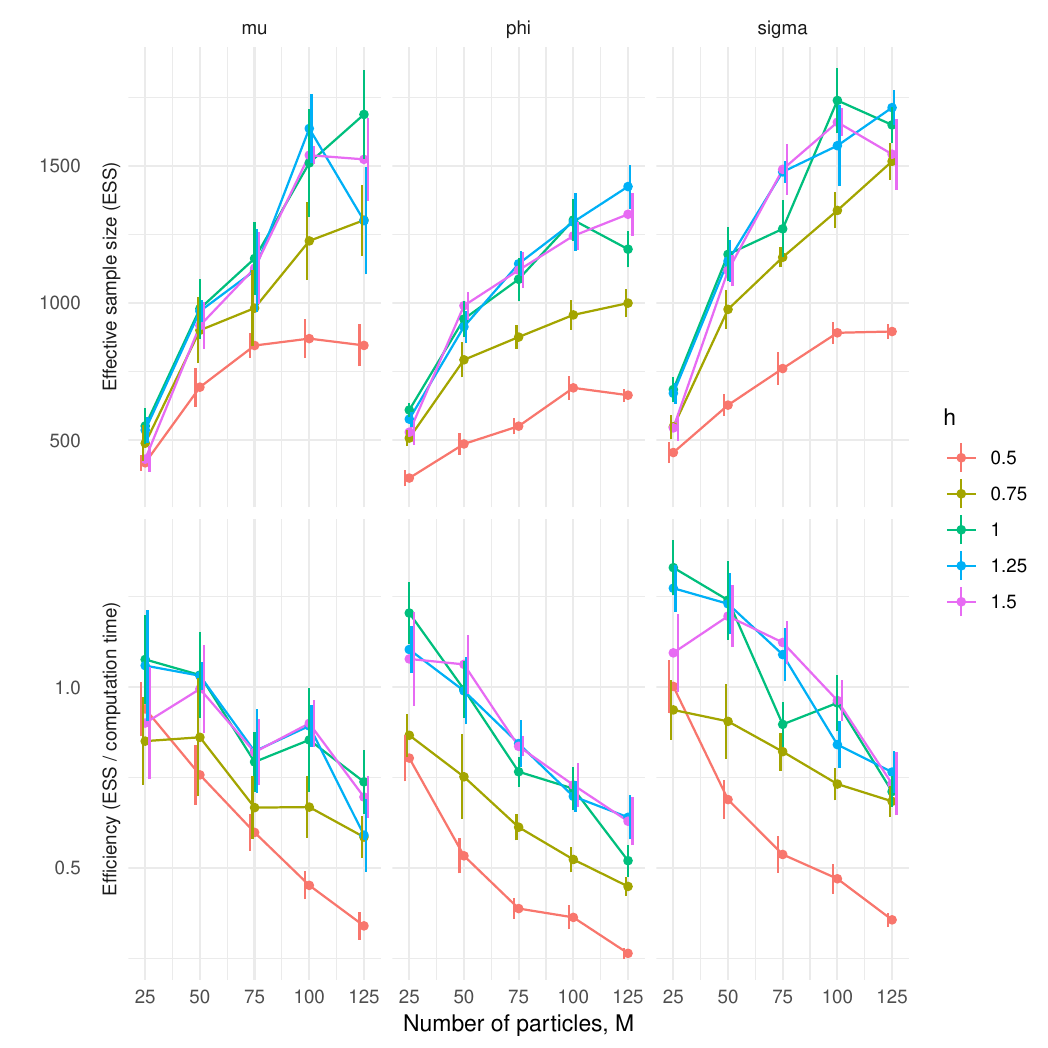}
  \caption{PMCMC effective sample size (ESS) and efficiency (ESS / computation time) for the stochastic volatility
  example for different numbers of particles (M) and proposal scales (h).}
  \label{fig:PMCMCplot}
\end{figure}

To illustrate these tradeoffs, we ran a set of computational experiments (Figure~\ref{fig:PMCMCplot}).  We compared  different values of SMC sample size, $M$, and scale of the multivariate normal random-walk proposals, $h$ (see below).  Values of $M$ were $25, 50, \ldots, 125$.  Values of $h$ were $0.5, 0.75, \ldots, 1.5$.  For each case (values of $M$ and $h$), results were compared using effective sample size (ESS) and efficiency, defined as ESS per computation time (in seconds).  Efficiency is the number of effectively independent samples generated per second.  To estimate ESS, we used the overlapping batch means (\code{obm}) method of package \pkg{mcmcse} \citep{flegal17}, with a batch size of 500.  While this is larger than necessary in most cases, there are some cases where the autocorrelation can be non-zero for lags of several hundred iterations, so the choice of 500 is conservative.  Ten replicates of each case were run.  Figure ~\ref{fig:PMCMCplot} shows mean +/- one standard error for ESS and efficiency for each case.

Proposals were multivariate normal with mean equal to the current values of $(\mu, \phi^{*}, \Omega)$ and diagonal covariance matrix with standard deviations ($0.89 h$, $0.39 h$, $1.45 h$).  These standard deviations are initial estimates (from a preliminary PMCMC run) of the posterior standard deviations multiplied by $h$.  This scheme provides a simply way to explore the role of proposal scale by using a single variable, $h$.  One could also experiment with using univariate proposals, which typically allow larger moves in one direction at a time but would require more SMC evaluations.  The setup of the present experiments provided a pragmatic way to gain insight on the impacts of proposal scale and SMC sample size on MCMC efficiency and to illustrate the tradeoffs involved.

While the results illustrate that estimates of ESS are noisy, clear patterns nevertheless emerge.   Increasing the number of particles improves mixing (ESS, Fig. \ref{fig:PMCMCplot}, top row) but is often not worth its computational cost (Fig. \ref{fig:PMCMCplot}, bottom row).  A doubling in the number of particles will roughly double the computation time, so it must at least double ESS to be worth the cost, and generally this is not the case in this example.   However, small numbers of particles demonstrate a pattern of ``stickiness'' that may not be satisfactory (Fig. \ref{fig:PMCMCtraceplot}, left panel).  Therefore, a good choice would be use to sufficient particles to avoid excessive stickiness while not imposing too great a computational cost.  

In typical problems, one will want to use a larger number of particles.  We were surprised at how small we could choose $M$ in this example and still obtain good results, but this is an example where the model fits the data reasonably.  In a case where the model cannot fit the data well, the variance in particle filter likelihood approximations will be higher, potentially causing stickiness and calling for more particles.

\subsection{IF2}
The steps to create an IF2 algorithm are similar to those for PMCMC: build a model, build the algorithm, compile both, and run.  For IF2, the example is a Gaussian random-walk with Gaussian measurement error of Nile river flows and a changepoint at a known time, following \citet{durbin12}, \pkg{dlm} and \pkg{FKF}.  

\subsubsection{Nile river flow example}
Define $x_t$ to be the state of flow in year $t$ and $y_t$ to be measured flow.  The model is:
  
\begin{align*}
  x_t &\sim N(I_{29}(t) c + x_{t-1}, \sigma^2)\\
  y_t &\sim N(x_t, \sigma_{M}^{2})\\
\end{align*}
where $I_{29}(t)$ is 1 when $t = 29$ and 0 otherwise, $c$ is the changepoint shift in mean flow, and $\sigma$ and $\sigma_M$ are standard deviations of the innovations and measurements, respectively.  (``Innovation'' refers to process stochasticity in state-space models.) Year 29 corresponds to 1899, when the Aswan Dam apparently changed Nile flows.  Package \pkg{dlm} implements this changepoint with an inflated variance of state dynamics.  We instead model this concept with a shift in the mean.  By use of maximum likelihood estimation, we found that the two approaches yield similar models and maximum likelihoods (not shown).  Maximum likelihood estimates obtained using package \pkg{FKF} are $\hat{\sigma}=0.01$, $\hat{\sigma}_M=127$, and $\hat{c}=-267$.  Thus, the model with MLE parameters has essentially no stochasticity in the latent states.  The maximum log likelihood is -626.4.

\proglang{BUGS} code for this model in \pkg{nimble} is: 
\begin{Schunk}
\begin{Sinput}
R> nileCode <- nimbleCode({
+   for(t in 1:n)
+     y[t] ~ dnorm(x[t], sd = sigmaMeasurements)
+   x[1] ~ dnorm(x0, sd = sigmaInnovations)    
+   for(t in 2:n)
+     x[t] ~ dnorm((t-1==28)*meanShift1899 + x[t-1], 
+                  sd = sigmaInnovations)
+   logSigmaInnovations ~ dnorm(0, sd = 100)       ## Prior is not used by IF2
+   logSigmaMeasurements ~ dnorm(0, sd = 100)      ## Prior is not used by IF2
+   sigmaInnovations <- exp(logSigmaInnovations)
+   sigmaMeasurements <- exp(logSigmaMeasurements)
+   x0 ~ dnorm(1120, var = 100)          ## Prior is not used by IF2
+   meanShift1899 ~ dnorm(0, sd = 100)   ## Prior is not used by IF2
+ })
\end{Sinput}
\end{Schunk}

The prior distributions are not used by IF2, except possibly to obtain boundaries of valid parameter values, but that is not the case here.

The next step is to build the model and algorithm and compile both:
\begin{Schunk}
\begin{Sinput}
R> y <- Nile
R> nileModel <- nimbleModel(nileCode, data = list(y = y),
+   constants = list(n = length(y)),
+   inits = list(logSigmaInnovations = log(sd(y)),
+   logSigmaMeasurements = log(sd(y)), meanShift1899 = -100))
R>                               
R> perturbThetaSD <- c(0.1, 0.1, 5)
R> initParamSigma <- c(0.1, 0.1, 5)
R> 
R> ff <- buildIteratedFilter2(model = nileModel,
+   nodes = 'x', params = c('logSigmaInnovations', 'logSigmaMeasurements',
+                            'meanShift1899'),
+   baselineNode = 'x0',
+   control = list(sigma=perturbThetaSD, initParamSigma = initParamSigma))
R> cNileModel <- compileNimble(nileModel)
R> cff <- compileNimble(ff, project = nileModel)
\end{Sinput}
\end{Schunk}
In \code{buildIteratedFilter2}, \code{nodes} gives the vector of latent state nodes, \code{params} gives the parameters to be optimized over, and \code{baselineNode} gives the node for the initial time point of the latent state (which should not have any data dependent on it), if applicable.  The \code{control} list elements include \code{sigma}, the vector of initial perturbation standard deviations ($\sigma_d$ values in Section~\ref{subsubsec:IF2}, in the order of \code{params}) to be multiplied by a cooling factor, and \code{initParamSigma}, the  vector of standard deviations of the initial particle swarm of parameters (also in the order of \code{params}).

\begin{Schunk}
\begin{Sinput}
R> numParticles <- 1000
R> numPFruns <- 100
R> alpha <- 0.2
R> 
R> est <- cff$run(m = numParticles, niter = numPFruns, alpha = alpha)
\end{Sinput}
\end{Schunk}

Here \code{m} is the number of particles, \code{niter} is the number of iterations of the IF2 algorithm, and \code{alpha} is the $\alpha$ in the equation for $c_{i, t}$  used to specify the cooling schedule via diagonal elements of $\Sigma_{i, t}$.

\subsection{Tuning IF2}
IF2 presents a user with numerous tuning parameters.  These include the number of particles, number of particle filter runs, cooling parameter $\alpha$, initial perturbation standard deviation for each parameter, and initial standard deviation for each parameter.  To explore and illustrate how choices of these parameters can affect performance, we ran a computational experiment with different choices for the number of particles and the cooling parameter.  Specifically, we ran the Nile flow example with \code{alpha} values of 0.1, 0.2, 0.4 and 0.6 and $M$ (number of particles) values of 100, 200, 500, 1000, and 2000.  Initial values were $\log(\mbox{sd}(y_{1:T}))$ for both log sigmas and -100 for the changepoint 
shift, $c$.  Perturbation standard deviations were 0.1, 0.1, and 5, and standard deviations of the initial particle swarm were the same.

\begin{figure}
  \includegraphics[width=6in]{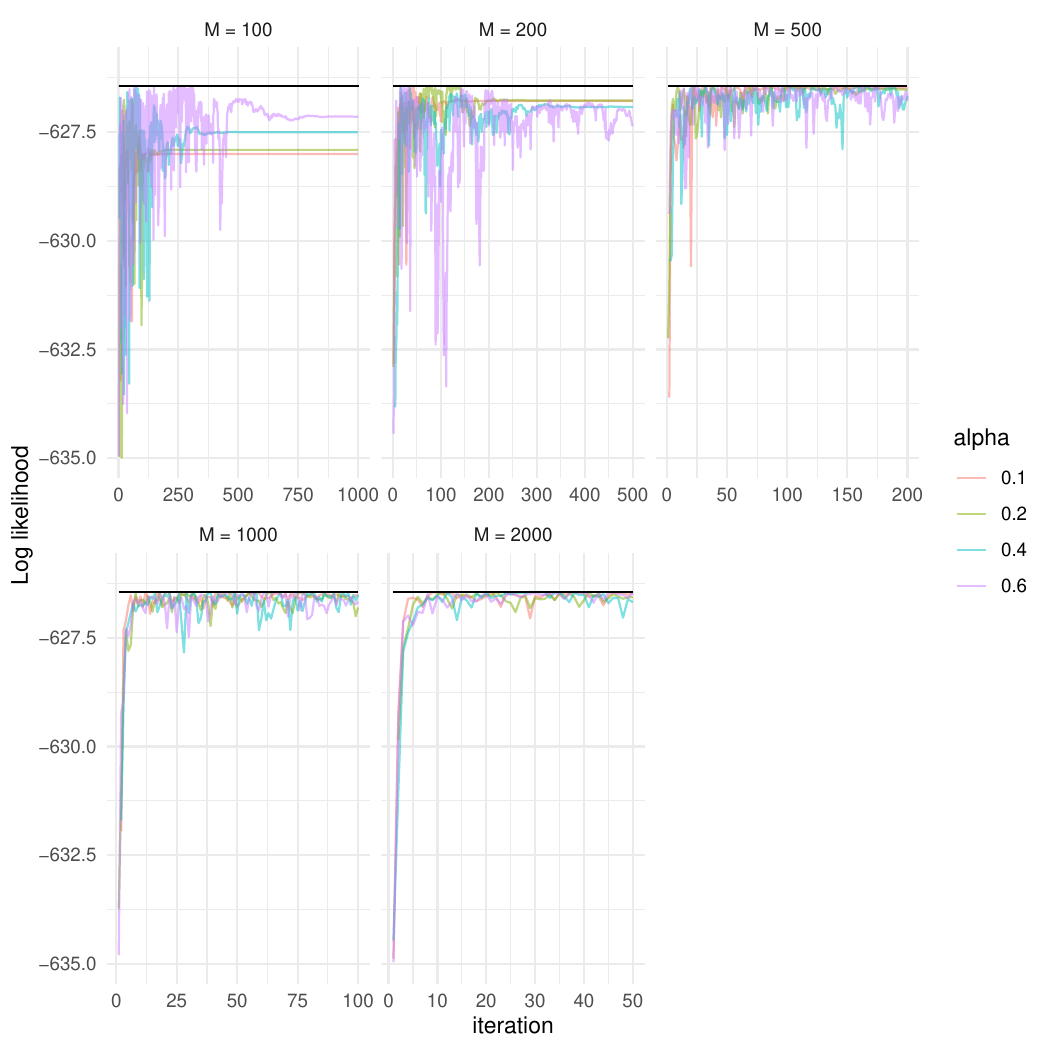}
  \caption{IF2 log likelihood vs. iteration for the Nile river flow example for
      different numbers of particles (M) and cooling parameters ($\alpha$).}
  \label{fig:IF2iteration}
\end{figure}

For each combination of $\alpha$ and $M$, we ran IF2 for a number of iterations such that there were $10^5$ total particles simulated through the entire time-series, or $10^7$ particle time-steps (since the length of the data is $T = 100$).  For example, with $M = 100$, we used $I = 1000$ iterations, while with $M = 500$, we used $I = 200$ iterations.  Since computational cost is closely related to number of particle time steps, this arrangement means that each case was run for a similar total computational effort.  The horizontal axis in figure~\ref{fig:IF2iteration} shows iterations but can be easily interpreted in terms of computational cost.  For example, half-way across the x-axis represents a different number of iterations but similar computational cost for each case.

To examine the path to convergence in each case, we calculated the correct likelihood using the Kalman Filter at each iteration of IF2.  The ability to calculate the correct likelihood  is the reason for using a linear, Gaussian example.  

Figure \ref{fig:IF2iteration} reveals several useful insights about IF2. With too few particles, IF2 can move too noisily to find the MLE before the cooling schedule effectively stops further exploration, as seen in the $M = 100$ and $M = 200$ cases.  This problem is exacerbated by smaller choices of $\alpha$ (faster cooling schedule).  On the other hand, choosing a larger $\alpha$ will incur more computational cost by requiring more iterations to converge, as seen in the $M = 500$ case, where all values of $\alpha$ appear to be converging to the MLE but larger ones are doing so more slowly. Finally, at large values of $M$, further increases in $M$ will yield little additional benefit.  What we have not explored here are choices for initial perturbation standard deviations ($\sigma_d$).  In general, relatively small values for these can work well because perturbations are applied at every time step of every particle filter run.

\section[Programming SMC algorithms in nimble]{Programming SMC algorithms in \pkg{nimble}}\label{sec:nimbleFuncs}
 
 In this section, we show how \pkg{nimble}'s SMC algorithms are implemented as \code{nimbleFunction}s, a high-level system for programming model-generic algorithms.  We will not cover every detail of programming in \pkg{nimble}, but rather we wish to show how algorithms are expressed compactly in high-level code that gets compiled via C++.  This will allow other programmers to quickly adapt our functions for their own needs.  For a more detailed discussion of \code{nimbleFunction} programming, see \citet{devalpine16} or Section IV of the \pkg{nimble} User Manual \citep{nimble-manual}.
 
 We demonstrate programming in \pkg{nimble} by providing code for a bootstrap filtering algorithm.  Note that the code shown below is simpler than the actual implementation of the bootstrap filter available in \pkg{nimble} through the \code{buildBootstrapFilter} function.  However, this demonstration code is indeed a fully functional bootstrap filter.  The bootstrap filter included in the \pkg{nimble} package simply has more customization options, and can handle somewhat more complicated models, than the demonstration algorithm provided here.
 
\pkg{nimble} programming uses two-stage evaluation defined by a \code{nimbleFunction}, which has two different types of code:  \code{setup} code and \code{run} code.  When a \code{nimbleFunction} is called, the \code{setup} code is evaluated first.   \code{setup} code is written in \proglang{R} and is primarily used to extract model information for later use in the \code{run} code.  The purpose of extracting model information is to specialize the algorithm to a particular model.  After \code{setup} code has been run once, the \code{run} code can be executed many times. \code{run} code is written in the \pkg{nimble} domain-specific language (DSL) for hierarchical model algorithms, which allows the code to be compiled into \proglang{C++}, in turn providing efficient execution of an algorithm's computations.  \code{run} code can make use of objects created in the \code{setup} code.
 
 The first \code{nimbleFunction} shown below, named \code{bootstrapFilter}, includes \code{setup} and \code{run} code that manage the time-iterations of the bootstrap filter.  It uses a list of other \code{nimbleFunction}s, each of which is responsible for one time step.  To set up this list, it iterates through time and calls the \code{nimbleFunction} \code{bootstrapStep} for each time.  The \code{setup} code of \code{bootstrapStep} takes information about the latent state at time $t$ and determines how to use the model to conduct Steps~\ref{line:bf-kloop}  through \ref{line:bootLikelihood} of the bootstrap filter algorithm given in Section~\ref{subsubsec:Boot} in its \code{run} code.  As the filtering algorithm progresses through each time point, samples from the filtering distribution at that time will be saved in \pkg{nimble} \code{modelValues} objects.  \code{modelValues} objects provide generic containers for storing sets of values of model nodes.
 
 Below is the call to \code{nimbleFunction} that defines \code{setup} and \code{run} code for the \code{bootstrapFilter}.  The \code{setup} code takes an argument called \code{model}, which must be a \pkg{nimble} model object created by a call to \code{nimbleModel} (described in Section~\ref{subsec:nimbleModels}).  It also takes the names of the latent states as argument \code{latentNodes}.  The \code{setup} code first specializes a \code{nimbleFunction} that will initialize the model and then obtains the names and dimensions of the latent states in the model in time order.  Two \code{modelValues} objects are created to store samples from the latent states.  The \code{mvWSamples} object will store a non-equally weighted sample, while \code{mvEWSamples} will store an equally weighted sample.  Finally, the \code{setup} code creates a list of \code{bootstrapStep} functions (called a \code{nimbleFunctionList}).  For each time point $t=1,\ldots,T$, the list contains one \code{bootstrapStep} function, which will conduct one time-step of the bootstrap filter.  We note that the creation of a separate \code{bootstrapStep} function for each time $t$ is necessary to allow the latent state $x_{t}$ at each time  to have arbitrary observation dependencies $y_{t}$, which may even have been declared with different model code for different times.
 
\begin{Schunk}
\begin{Sinput}
R> bootstrapFilter <- nimbleFunction(
+  setup = function(model, latentNodes) {
+    my_initializeModel <- initializeModel(model)
+    latentNodes <- model$expandNodeNames(latentNodes, sort = TRUE)
+    dims <- lapply(latentNodes, function(n) nimDim(model[[n]]))
+    mvWSpec <- modelValuesConf(vars = c('x', 'wts'),
+                               types = c('double', 'double'),
+                               sizes = list(x = dims[[1]], wts = 1))
+    mvWSamples <- modelValues(mvWSpec)
+    mvEWSpec <- modelValuesConf(vars = c('x'), types = c('double'),
+                                sizes = list(x = dims[[1]]))
+    mvEWSamples <- modelValues(mvEWSpec)
+    bootStepFunctions <- nimbleFunctionList(bootstrapStepVirtual)
+    timePoints <- length(latentNodes)
+    for (t in 1:timePoints)
+      bootStepFunctions[[t]] <- bootstrapStep(model, mvWSamples,
+                                              mvEWSamples, latentNodes, t)
+  },
+  run = function(M = integer()) {
+    my_initializeModel$run()
+    resize(mvWSamples, M)
+    resize(mvEWSamples, M)
+    for (t in 1:timePoints)
+      bootStepFunctions[[t]]$run(M)
+  }
+ )
\end{Sinput}
\end{Schunk}
 
 The \code{run} code for the \code{bootstrapFilter} function takes as its only argument the number of particles ($M$) to use for estimation. \code{run} code requires explicit declaration of the type of any arguments, so here $M$ is specified as a scalar integer.  In general, the type of a return object  must also be declared, although this function does not return anything so no declaration is necessary.  The \code{run} function first initializes the model (conducting Steps~\ref{line:init1} and \ref{line:init2} of the bootstrap filter algorithm), and then re-sizes the \code{modelValues} objects so that they can store $M$ particles.  After that, the \code{run} function iterates through each time point, running the \code{bootstrapStep} function that was defined for that time point in the \code{setup} code.  To keep the example simple, this version does not provide an estimate of the likelihood $\tilde{p}(y_{1:T})$.
  
Creating a \code{nimbleFunctionList}, such as the one used in the \code{setup} code above, requires an additional piece of code that informs \pkg{nimble} about the input arguments and return objects of each function in that list.  Specifically, the \code{nimbleFunctionVirtual} function is used to define the methods and their argument and return types that each element in the \code{nimbleFunctionList} will have.  Below, we specify that each element of our \code{nimbleFunctionList} will have a run function with a single integer input and no return object.
 
\begin{Schunk}
\begin{Sinput}
R> bootstrapStepVirtual <- nimbleFunctionVirtual(
+  run = function(M = integer()) {}
+ )
\end{Sinput}
\end{Schunk}
 
 \code{setup} and \code{run} code for the \code{bootstrapStep} function are given below.  At each time point $t$, the \code{setup} function gets the names and deterministic dependencies of the previous and current latent states.  The \code{run} code first declares a length $M$ vector of integers (to store particle indices) and a length $M$ vector of doubles (to store particle weights).  The \code{run} code then iterates through the particles.  For each particle, the code takes the value of the latent state at $t-1$ from the equally weighted \code{modelValues} object, uses that value to propagate a value for the latent state at time $t$, and calculates a weight.  The particles and corresponding weights are stored in the non-equally weighted \code{modelValues} object.  Finally, particles are resampled proportional to their weights and the resampled particles are stored in the equally weighted  \code{modelValues} object.
 
 In this algorithm particles are propagated using the proposal distribution $q({x}_{t}|x_{t-1}^{(m)},y_{t})=f({x}_{t}|x_{t-1}^{(m)})$, which simplifies the weight calculation in Step~\ref{line:bootWeight} of Algorithm~\ref{bootstrap_algo}.  Additionally, since resampling is performed at each time point, weights from time $t-1$ do not need to be used when calculating weights at time $t$.  This results in a weight calculation of $w_{t}^{(m)}=g(y_{t}|\tilde{x}_{t}^{(m)})$. 
 
\begin{Schunk}
\begin{Sinput}
R> bootstrapStep <- nimbleFunction(
+  contains = bootstrapStepVirtual,
+  setup = function(model, mvWSamples, mvEWSamples, latentNodes,
+                   timePoint) {
+    notFirst <- timePoint != 1
+    prevNode <- latentNodes[if(notFirst) timePoint - 1 else timePoint]
+    thisNode <- latentNodes[timePoint]
+    prevDeterm <- model$getDependencies(prevNode, determOnly = TRUE)
+    thisDeterm <- model$getDependencies(thisNode, determOnly = TRUE)
+    thisData   <- model$getDependencies(thisNode, dataOnly = TRUE)
+  },
+  run = function(M = integer()) {
+    ids <- integer(M, 0)
+    wts <- numeric(M, 0)
+    for(m in 1:M) {
+      if(notFirst) {
+        copy(from = mvEWSamples, to = model, nodes = 'x',
+             nodesTo = prevNode, row = m)
+        model$calculate(prevDeterm)
+      }
+      model$simulate(thisNode)
+      copy(from = model, to = mvWSamples, nodes = thisNode,
+           nodesTo = 'x', row = m)
+      model$calculate(thisDeterm)
+      wts[m] <- exp(model$calculate(thisData))
+      mvWSamples['wts', m][1] <<- wts[m]
+    }
+    rankSample(wts, M, ids)
+    for(m in 1:M){
+      copy(from = mvWSamples, to = mvEWSamples, nodes = 'x',
+           nodesTo = 'x', row = ids[m], rowTo = m)
+    }
+  })
\end{Sinput}
\end{Schunk}

 The calls to \code{calculate} within the above \code{run} code serve two purposes.  The first two \code{calculate} calls are used to calculate the values of any deterministic dependencies of the latent state, as these dependencies must be recalculated any time the latent state takes on a new value.  The third call to \code{calculate} is used to calculate the log-likelihood of the data given the current latent state value, which is then used as a particle weight.   The \code{rankSample} function fills the elements of the \code{ids} vector with the indices of the particles that have been chosen in the resampling procedure.
 
 Once the \code{nimbleFunction}s have been defined, we can build, compile, and run the bootstrap filter. The code below runs the example filter on the \code{exampleModel} of Section~\ref{subsec:fixedParam} and creates a histogram of samples from the filtering distribution of $x$ at the last time point.

\begin{Schunk}
\begin{Sinput}
R> myBootstrap <- bootstrapFilter(exampleModel, 'x')
R> cmyBootstrap <- compileNimble(myBootstrap, project = exampleModel,
+   resetFunctions = TRUE)
R> cmyBootstrap$run(1000)
R> filterSamps <- as.matrix(cmyBootstrap$mvEWSamples, 'x')
R> hist(filterSamps, main = '', xlab = '')
\end{Sinput}
\begin{figure}[H]
\includegraphics[width=\maxwidth]{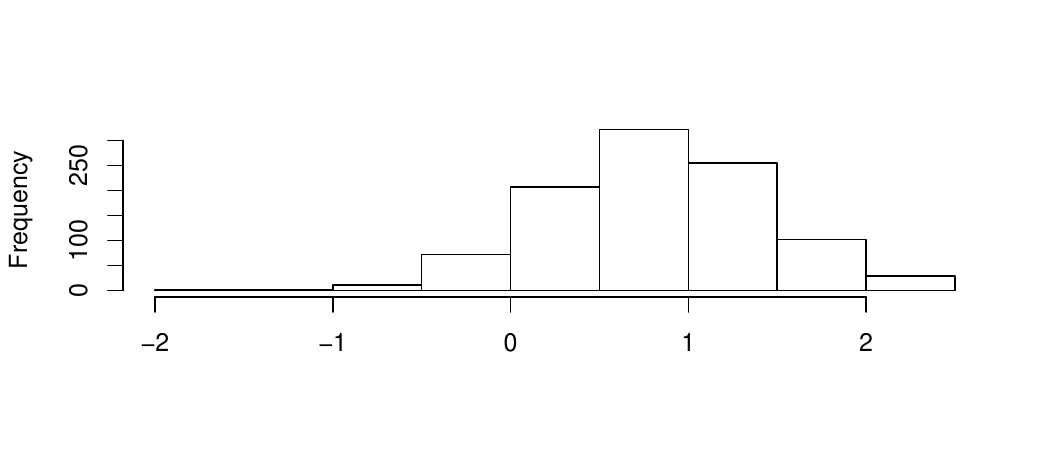} \caption[A histogram of the filtering distribution of $x$]{A histogram of the filtering distribution of $x$.}\label{fig:bootRun_Chunk}
\end{figure}
\end{Schunk}
 
The bootstrap filter code provided above demonstrates \pkg{nimble}'s ability to program model-generic algorithms.  The filter could be used to conduct filtering on any correctly specified state-space model.  In addition to the generality of the algorithm, it would be relatively straightforward to modify the filter, changing it to an auxiliary particle filter, an IF2 algorithm, or a filter type not currently included in \pkg{nimble}.  The ease with which existing algorithms can be modified, along with the generality with which they are written, promotes the development of user-written filters.
 
\section{Conclusion}

This paper has described \pkg{nimble}'s suite of SMC algorithms, which provide straightforward methods of conducting inference on state-space models.  In addition, \pkg{nimble}'s model-generic programmability make it well-suited for implementing new SMC algorithms, an example of which was given in Section~\ref{sec:nimbleFuncs}. \pkg{nimble}'s  flexible model specification also enables the application of existing algorithms to more general models or new settings.  For example, a model could be written where a number of state-space models are set within a larger hierarchical structure.  Using \pkg{nimble}, SMC algorithms could be used to estimate the individual state-space models, while an MCMC algorithm could conduct inference on higher-level parameters.   \code{nimble} also provides easily accessible model comparison tools that can be used in conjunction with its state-space modeling algorithms, which we hope will allow users to answer previously difficult questions about their time-series data.
 
Additional examples of modeling and inference using \pkg{nimble} can be found at \url{https://r-nimble.org}.  

\section*{Acknowledgements}

This work was supported by the U.S. National Science Foundation under grants DBI-1147230 and ACI-1550488, and by the Google Summer of Code.  

\bibliography{jss_citations}
 
\end{document}